\documentclass[useAMS,usenatbib]{mn2e}

\usepackage{graphicx}
\usepackage{amssymb,amsmath}
\usepackage{booktabs}
\usepackage{afterpage}
\usepackage{pdflscape}

\title[Stellar density maps of the Northern Galactic Plane]{Calibrated and completeness-corrected optical stellar density maps of the Northern Galactic Plane}
\author[H. J. Farnhill et al.]{H. J. Farnhill$^{1}$\thanks{E-mail:
h.farnhill@herts.ac.uk}, J. E. Drew$^1$, G. Barentsen$^{1,2}$, E.A. Gonz\'{a}lez-Solares$^3$ 
\\
$^{1}$School of Physics, Astronomy \& Mathematics, University of Hertfordshire, College Lane, Hatfield, AL10 9AB, U.K.\\
$^{2}$NASA Ames Research Center, Mail Stop 245-30, Moffett Field, CA 94035, USA\\
$^{3}$Institute of Astronomy, University of Cambridge, Madingley Road, Cambridge CB3 OHA, UK}
\begin{document}

\date{Draft last compiled \today}

\pagerange{\pageref{firstpage}--\pageref{lastpage}} \pubyear{2015}

\maketitle

\label{firstpage}

\begin{abstract}
Following on 
from the second release of calibrated photometry from IPHAS, the INT/WFC Photometric H$\alpha$ 
Survey of the Northern Galactic Plane, we present incompleteness-corrected stellar density maps 
in the $r$ and $i$ photometric bands.  These have been computed to a range of limiting magnitudes 
reaching to 20th magnitude in $r$ and 19th in $i$ (Vega system), and with different angular 
resolutions -- the highest resolution available being 1 square arcminute.  The maps obtained cover 
94 percent of the 1800 square-degree IPHAS footprint, spanning the Galactic latitude range, 
$-5^{\circ} < b < +5^{\circ}$, north of the celestial equator.  The corrections for incompleteness, 
due to confusion and sensitivity loss at the faint limit, have been deduced by the method of 
artificial source injection. The presentation of this method is preceded by a discussion
of other more approximate methods of determining completeness.  
Our method takes full account of position-dependent seeing and source ellipticity in the 
survey database.  The application of the star counts to testing reddened Galactic disc models
is previewed by a comparison with predicted counts along three constant-longitude cuts 
at $\ell \simeq 30^{\circ}$, $90^{\circ}$ and $175^{\circ}$: some over-prediction of 
the most heavily reddened $\ell \simeq 30^{\circ}$ counts is found, alongside good agreement at 
$\ell \simeq 90^{\circ}$ and $175^{\circ}$.

\end{abstract}


\section{Introduction}

The positioning of the Solar System almost in the equatorial plane of the Milky
Way places a significant obstacle in the way of understanding the structure of our 
own galaxy.  Despite the fact that the disc of the Milky Way is effectively 
the largest object in the night sky, offering vastly superior angular resolution 
compared to that achievable for any other galaxy, sightlines at low Galactic latitude 
remain a challenge to decipher because of large and variable amounts of dust 
extinction.  Given that the formation and maintenance of galactic 
discs is an important problem in galaxy evolution \citep{vanderKruit2011}, an improved 
vision of our own galactic disc is needed.  And as our home in the Universe, it is of 
interest in its own right.

The reliable empirical determination of the 3-dimensional (3D) distribution of the Milky 
Way's interstellar dust, needed to make sense of the disc, is now becoming possible 
through increasingly sophisticated analyses of comprehensive digital survey data 
\citep{Drimmel2001,Marshall2006,Sale2009,Sale2014}.  In the
next 5-10 years, these advances will complement the astrometric harvest being gathered 
by the Gaia mission and usher in a much better, sharper vision of the 3D Milky Way.
Nevertheless, the anticipated Gaia parallax precision at fainter magnitudes ($100\mu$arcsec 
at $G \sim 19$) will still leave stars beyond the first 1--2 kpc with distances known to 
a precision no better than $\sim$10-20 percent.  Accordingly, it remains useful to supplement 
our knowledge through the application of other methods that can test predictive Galactic 
models.  One of these, that puts to good use the uniquely detailed view we have of the 
Galactic disc, is comparison with magnitude-limited star counts.  This approach has 
been applied successfully in the past and continues to be influential in guiding the 
content of Galactic models \citep{Robin2003, Czekaj2014}.  

So far, deeper optical star-count mapping has only been carried out in the 
southern Galactic plane \citep{Ruphy1997}.  The options to conduct a 
well-calibrated stellar density mapping of the northern sky are now 
appearing in the wake of data product releases from wide-area digital 
imaging surveys \citep{Stoughton2002,Magnier2013}.  
The greatest challenge is at low Galactic latitudes, but even this is now 
becoming tractable as the 1 arcsec angular-resolution IPHAS \citep{Drew2005}, and 
UVEX \citep{Groot2009} surveys approach completion.  In particular, the 
recent release of IPHAS DR2 \citep{Barentsen2014}, offering uniform 
photometry of the northern Galactic Plane in $r$, $i$ (and 
$H\alpha$), provides the basis for a precise and deep stellar density map 
across almost 1800 square degrees.

In this work, we describe the construction of $r$, $i$ stellar density maps at
a range of angular resolutions, up to a maximum of 1 square arcminute, 
based on IPHAS DR2.  In order that this mapping is not hampered by variable 
observing conditions and correspondingly variable levels of source loss, great care has been 
taken to make corrections for incompleteness by evaluating the 
results of artificial source injection tailored to every survey field. In the Galactic 
Plane where confusion can present as a significant issue, it is important to do this 
reliably.  Because literature presentations of source injection and other methods 
of correction have so far been limited, (but see e.g. \citet{Harvey2006} for more 
than usual detail), we present a reasonably full description of how we arrive at 
the position- and magnitude-dependent corrections applied.  This is presented in 
sections 3.3 to 3.7, after a brief restatement of the main features of the IPHAS 
survey in section 2 and some appraisal via two more approximate techniques in 
sections3.1 and 3.2.

The resulting corrected $r$, $i$ stellar density maps are described and briefly discussed
in section 4.  The maps themselves, with maximum magnitude limits of $r = 19$ and 
$i = 18$ respectively (Vega system), are presented as plots within an Appendix and 
are provided within machine-readable online supplementary material.   In section 5, a 
first purely-illustrative 
comparison is made between counts derived from the $i < 18$ map and Galactic model 
predictions.  This uses two distinct reconstructions of the 3D distribution of interstellar 
extinction in order to gain a first impression of how influential the choice of 
extinction map can be.  Interestingly, the results from this preliminary exploration are 
already mixed: we find that at $\ell \simeq 90^{\circ}$ and $175^{\circ}$, the agreement 
between the star counts and model prediction is good -- but at the lowest Galactic 
longitude examined, $\ell \simeq 30^{\circ}$, a significant discrepancy appears.  

The paper ends in section 6 with some concluding remarks, including comment on the best use
of these star-count maps.

\section[]{Observations}

\subsection{IPHAS broad-band photometry}

The most recent release, DR2, of the INT/WFC Photometric H$\alpha$ Survey of the Northern Galactic Plane (IPHAS) was 
presented by \citet{Barentsen2014}, while the basic specification of IPHAS was set down by \citet{Drew2005}.  This 
survey, conducted using the Wide Field Camera (WFC) mounted on the Isaac Newton Telescope (INT) in La Palma, 
provides photometry of the complete Northern Galactic Plane within the latitude range $-5^{\circ} < b < +5^{\circ}$, in narrow-band H$\alpha$ and broadband Sloan $r$ and $i$.  It is the broadbands that are the focus of this paper.  The typical $5\sigma$ magnitude limit reached in the survey is 21.2 in $r$ and 20.0 in $i$, achieved at a median seeing of 1.1 arcsec.  The typical external photometric precision of DR2 is close to 0.03 magnitudes, as judged by comparisons with SDSS DR9 data.  In Galactic longitude, this survey spans the range 
$30^{\circ} < \ell < 215^{\circ}$.  In time it will be fully complemented by the VPHAS+ survey \citep{Drew2014} covering the Southern Galactic Plane and Bulge.  

A highly characteristic feature of the IPHAS survey, due to its targeting the Galactic Plane, is that the stars dominating the faint end of the captured $r$ and $i$ band magnitude distributions are typically either intrinsically red or highly reddened.  Consequently, a reasonable generalisation to apply is that at $r = 19$, stars captured by the survey are typically brighter in $i$ by around one magnitude (in the Vega system, adopted by the survey -- see figs. 15 to 17 in 
\citet{Barentsen2014}).  Because of this, in the presentation that follows, comparisons between the $r$ and $i$ density maps obtained are generally drawn between those derived for $r = 19$ an $i = 18$ limits. 

Whilst IPHAS is certainly uniform in its execution, the weather at the telescope represented an important 
variable.  The observations were obtained via standard allocations of time, with the result that a wide variety 
of observing conditions are contained within the survey database.  Thanks to the opportunity to obtain repeats 
of fields exposed in poor conditions, it was possible to apply quality cuts in the preparation of IPHAS DR2 so 
as to omit clearly inferior data, whilst still achieving over 92\% coverage of the survey footprint.  Even so, 
there is still a broad quality range within DR2 in terms of both measured widths of the point spread function (PSF), and limiting magnitude \citep[for full details, see][]{Barentsen2014}.  This variation, along with over a 
factor of ten contrast in the typical observed stellar density along the Northern Plane 
\citep[see Fig.~3 in][]{Gonzalez-Solares2008}, makes it important to achieve uniformity of outcome through careful 
position-dependent completeness corrections. 
This task, along with the development and application of an algorithm that ensures reliable source 
counting without duplication and proper accounting for small gaps and other irregularities in the data, is described here (see also Farnhill 2015).

The first step in this process is to establish a working definition of a detected 'star' in the survey data.

\subsection{On morphological classification and the expected number density of extra-galactic sources}
\label{subsec:galaxy_density}

\begin{figure}
\begin{center}
\includegraphics[width=1\linewidth]{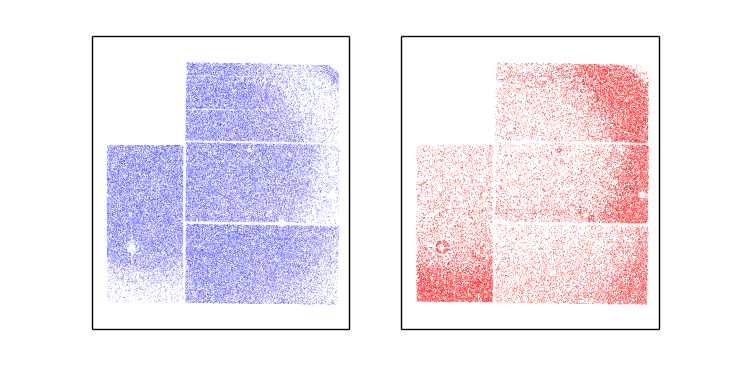} 
\caption{Distributions of sources classified as stellar ($rClass=-1$, blue) 
and non-stellar ($rClass=+1$, red) in the $r$-band for IPHAS field 4975o\_aug2004a.}
\label{fig:misclassification}
\end{center}
\end{figure}

A fraction of sources detected in DR2 fields are known to be misclassified as 
non-stellar towards field edges - this is due to progressive distortion in the PSF with increasing 
distance from the optical axis (OA) of the INT/WFC system. This pattern of behaviour has a small 
impact on the final DR2 catalogue, which provides primary detections that are selected according to 
a criterion of lowest distance to the OA.  But when counting stellar sources (morphological classes -1 
and -2) from a single field to produce a stellar density map the issue of progressive misclassification 
has to be tackled. Fig.~\ref{fig:misclassification} shows an example of a field exhibiting this problem.

Assuming a constant ratio of stellar to non-stellar sources applies in reality across 
any one field, comparisons between 5 sq. deg. regions at field centres and edges revealed
occasional misclassification of several hundred sources. Of course, some of these objects 
are not misclassified -- they can be genuine examples of galaxies presenting as extended objects.  
To determine the expected number of galaxies detected in a similar area, galaxy counts were 
taken from \citet{Yasuda2001}, down to the 5$\sigma$ $r$- and $i$-band limits 
of IPHAS, and then extinguished by the median \citet{Schlegel1998} extinction values for all fields (corrected using the \citet{Schlafly2011} recalibration). Based on median limiting magnitudes, 
$\sim 1$ extragalactic source is predicted per 5 sq. deg. region. Against stellar densities typically 
orders of magnitude larger, this is very small.  It was therefore regarded as safe to include 
'non-stellar' +1 classified sources for density mapping purposes - the losses that would be 
suffered on leaving them out, in fields similar to that depicted in Fig.~\ref{fig:misclassification}, 
are much larger than is the likely contribution from galaxies.

\section[]{Completeness correction} 
\label{sec:completeness}

Genuine astronomical sources falling within IPHAS detection limits can fail to
appear in the resulting photometric catalogues for a number of reasons.   

Detector issues can prevent sources from being picked up: the WFC is a 4-CCD mosaic leaving 
gaps between the component detectors, and there are also bad columns and regions of
vignetting that will hinder detection. In the majority of such cases (but not quite all),
missing sources will be picked up in the offset partner pointing, or in the overlap with a 
neighbouring field.

More significant and pervasive losses affecting the final star counts are those
due either to confusion promoted by high stellar densities or to statistical sensitivity losses at the 
faint limit for detection.  In the Galactic plane at lower longitudes, where the stellar density in 
IPHAS can reach to more than 100,000 per square degree, confusion will be especially important and liable to 
determine the effective magnitude limit.  Outside the Solar Circle where the typical stellar density is 
around 20,000 per square degree, confusion has only marginal significance: in this domain the median 1.1 
arcsec point spread function (PSF) implies 70 'beams' per source, to be compared with the rule of thumb 
confusion threshold of 30 per source (see \citet{Hogg2001}).   Both kinds of incompleteness are exacerbated by 
relatively poor seeing.  We first consider approximate methods of correction, to gain insight into
the role source confusion is playing within the IPHAS catalogues.  We then go on to describe our chosen 
method of incompleteness correction based on artificial source injection.

\subsection{Confusion, as estimated from nearest-neighbour analysis}

Confusion, the effect of background noise caused by unresolved sources, scales with source 
density \citep{Condon1974}. 
For randomly distributed sources, the probability of finding a given number of neighbours within angular 
distance $\theta$ can be described by a Poisson distribution, which leads to the nearest neighbour distribution 
\begin{equation}
n(\theta) = 2\rho^2\Omega\pi\theta\mathrm{e}^{-\rho\pi\theta^2}
\label{eq:theoretical_neighbours}
\end{equation}
\noindent where $\rho$ is the  density of stars per unit solid angle, and $\Omega$ is the sky solid angle of interest, as presented in \citet{Bahcall1983}.

\begin{figure}
\begin{center}
\includegraphics[width=1\linewidth]{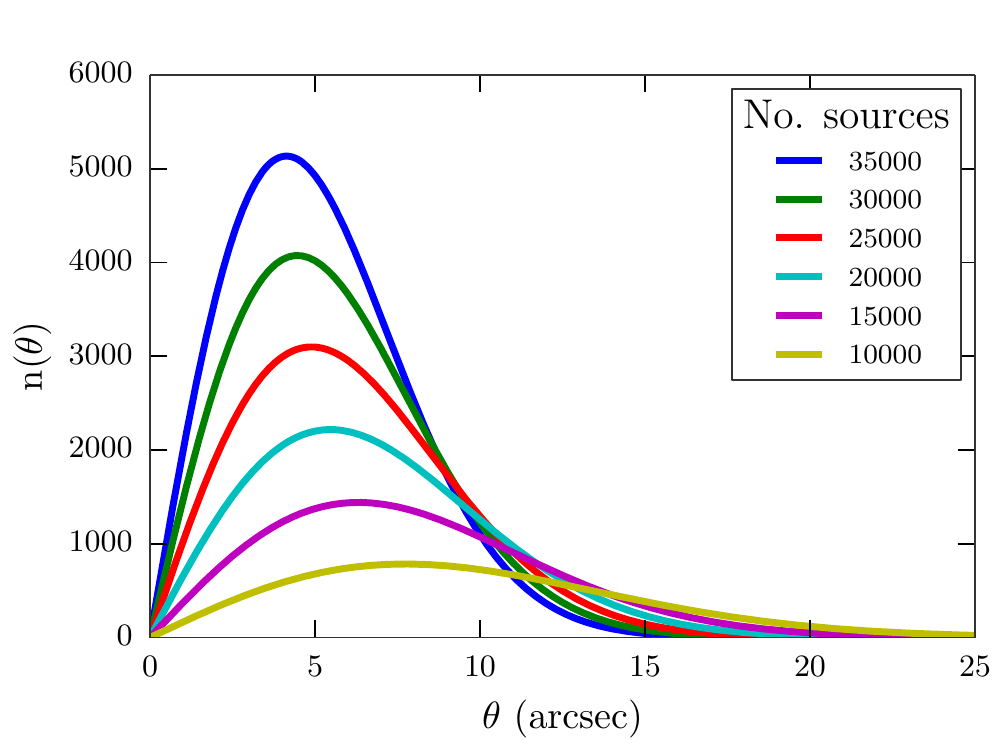} 
\caption{Theoretical distribution (see Equation  \ref{eq:theoretical_neighbours}) of nearest neighbour distances for an area with the effective area of a WFC image, for fields containing varying numbers of sources.}
\label{fig:theoretical_neighbours}
\end{center}
\end{figure}

Fig.~\ref{fig:theoretical_neighbours} shows this distribution for a sky area matching that of the WFC field 
of view (0.29 sq. deg.) containing different numbers of sources. It can be seen that increasing the density of sources 
increases the number of nearest neighbours at small separations, while lowering the number at greater separations,
thereby pulling in the peak separation value, $\theta_{max}$.  Equation~\ref{eq:theoretical_neighbours} holds for a 
population of sources randomly distributed regardless of brightness. A posteriori checks have revealed that, at the 
$\sim$1 arcsec resolution characteristic of IPHAS, the angular distribution of stars in the northern Galactic plane 
conforms quite well to the random case: essentially we find that the $Rn$ statistic or the ratio of observed mean nearest-neighbour separation to the predicted (random) value is commonly between 1 and 1.1, per field.  If clustering were significant, or indeed 
close pairs due to stellar multiplicity were frequent, $Rn$ would be pulled down well below unity.  In this context it is worth noting the following feature of IPHAS star counts: the magnitude range from $\sim13$th to 20th begins to sample the Galactic Plane from distances of 0.5--1~kpc, at the bright end, out to 4 or more kpc depending on sightline.  Accordingly IPHAS star counts say probe the Galactic disk beyond the first kiloparsec, rather than the solar neighbourhood.    

In order to reproduce the theoretical distribution (Eqn. \ref{eq:theoretical_neighbours}), all sources in 
the area $\Omega$ would need to be recovered. This will never be the case in dense fields, which will modify the 
observed distribution in the sense that the smallest separations will be under-reported - appearing to relatively
boost the proportion of nearest neighbours at larger values of $\theta$.

\begin{figure*}
\begin{center}
\includegraphics[width=1\textwidth]{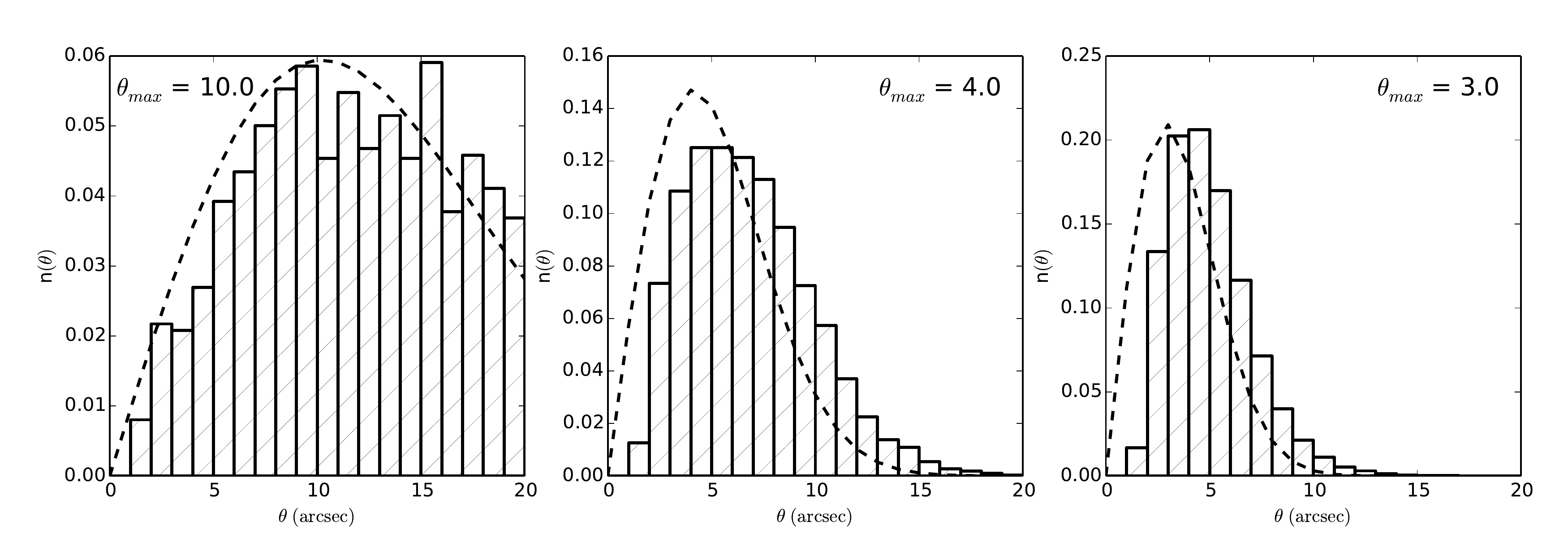} 
\caption{Nearest neighbour distributions for three IPHAS fields, 
normalised to allow the overplotting of theoretical distribution (dashed line).
 Density of fields increase from left to right.}
\label{fig:iphasfield_neighbours}
\end{center}
\end{figure*}

Three test cases are shown in Fig.~\ref{fig:iphasfield_neighbours}: these
IPHAS fields, located around $\ell\sim32^{\circ}$, were chosen because they lie in a
region of the survey containing fields spanning a wide range of densities. The
$n(\theta)$ distributions were generated by selecting subregions of the CCDs
such that a border of width 30$^{\prime\prime}$ was excluded -- the nearest
neighbour would then be identified for each source within the central subregion. 
This approach eliminates from consideration sources close to CCD edges whose
nearest neighbours are likely to be located in the image plane beyond the 
detector edge, whilst it rightly permits the nearest-neighbour search for each 
central-region source to extend into the borders. The border width was 
set at $30^{\prime\prime}$ after obtaining initial nearest neighbour 
distributions which showed that the great majority of nearest neighbours occurred 
at smaller separations than this (see e.g. Fig. 2).

The fields are displayed in order of increasing density. The left-most field,
shown contains 3,966 sources - a number sufficiently low that confusion would 
not be expected to make a significant contribution to incompleteness. The fields
presented in the middle and right panels, containing 24,291 and 49,150 sources 
respectively, suffer from increasing levels of confusion. The overplotted theoretical 
distribution gives an idea of the amount of confusion; at low field densities, the empirical and 
theoretical distributions agree quite well, while at higher densities the theoretical 
distributions clearly predict closer nearest neighbours than measured.

While the observed nearest neighbour distribution is affected by
confusion most significantly at smaller ($\lesssim10^{\prime\prime}$) separations, the 
observed number of nearest neighbours at larger separations will be less affected. 
Hence, the theoretical distribution best fitting the tail at large $\theta$ values comes 
close to describing the field as if confusion were not an issue.  We have exploited
this property as a means to gauge confusion loss at small separations.  We have fitted
distributions generated from Eqn. \ref{eq:theoretical_neighbours} to the observed 
distributions at $>10^{\prime\prime}$ separation, varying $\rho$ such that the number of 
sources in a field ranged from half to twice its observed value. By calculating $\chi^2$ 
for each fit at large separations, the best fitting value of $\rho$ was found for each, 
identifying the correction that should be applied for confusion.

\begin{figure}
\begin{center}
\includegraphics[width=1\linewidth]{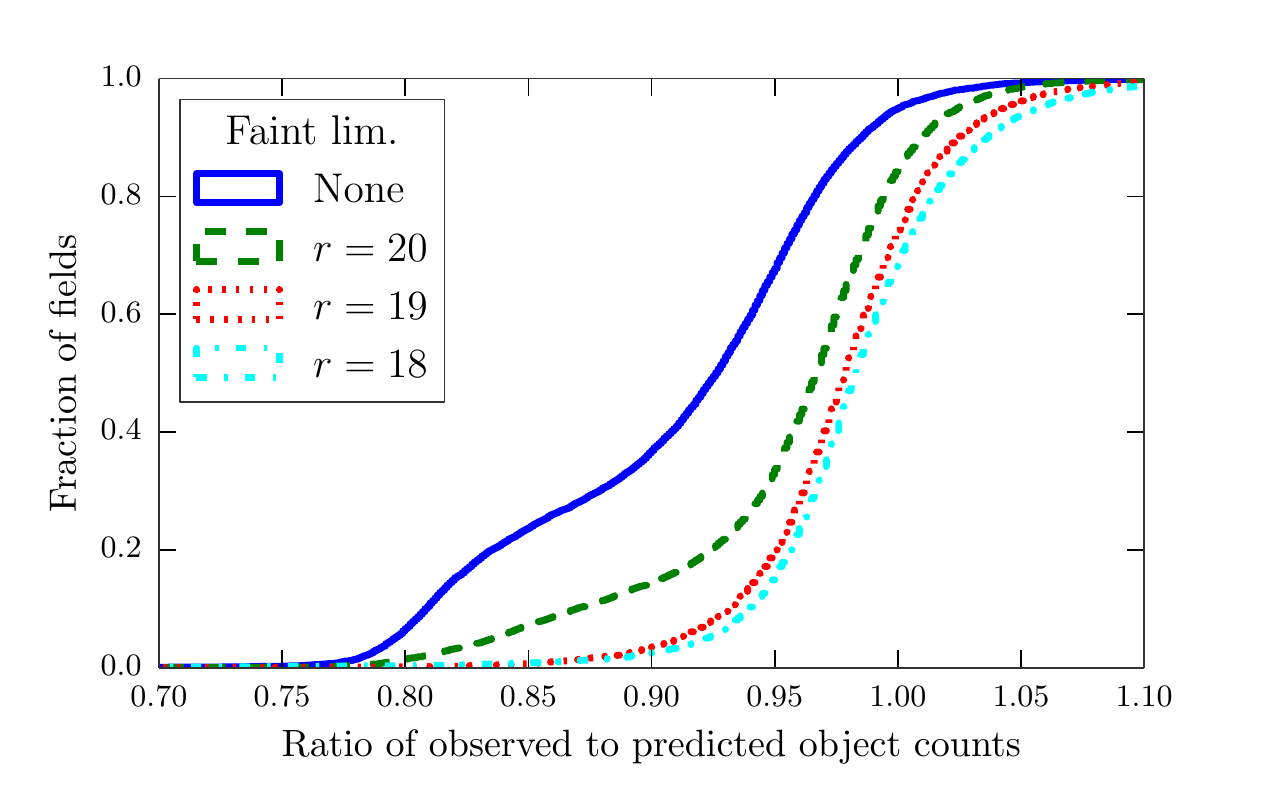} 
\includegraphics[width=1\linewidth]{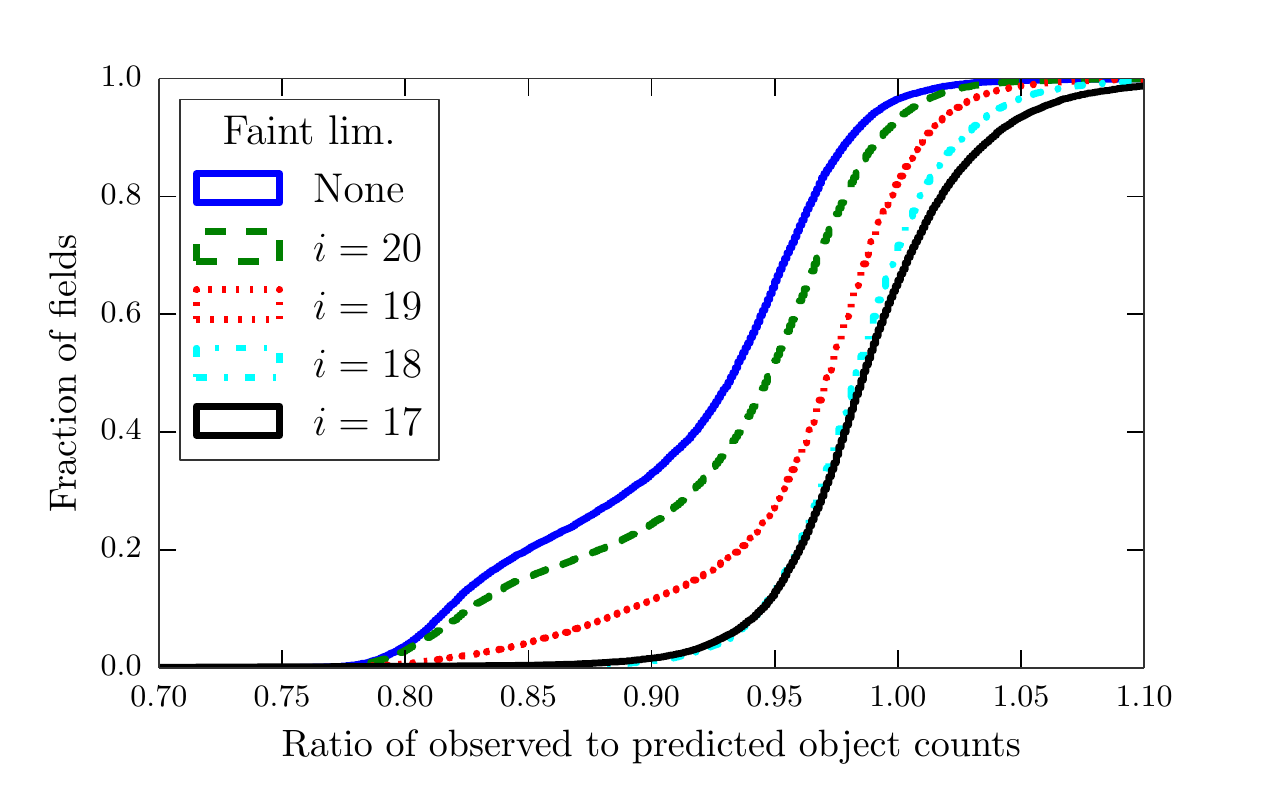} 
\caption{Cumulative histogram of $\dfrac{observed}{predicted}$ 
source counts in $r$ (upper panel) and $i$ (lower panel), where the predicted 
counts were estimated by fitting Equation \ref{eq:theoretical_neighbours} to 
the tail of the nearest neighbour distribution. The curves in both panels 
represent the ratio as determined for sources down to different limiting 
magnitudes, plus an instance where no limit was placed on the magnitude 
of sources.}
\label{fig:confusion_completeness}
\end{center}
\end{figure}

Fig.~\ref{fig:confusion_completeness} shows the effect of confusion on the
survey as a whole. The best fitting n$(\theta)$ distribution (corresponding to a
theoretical unconfused source count) was obtained for each IPHAS field, allowing a 
picture to be built up of the impact of confusion. The statistic chosen to 
represent the magnitude of the correction is the ratio of $\frac{observed}{predicted}$ 
sources, where the ``predicted" source count is the count corresponding to the 
best fit theoretical n$(\theta)$ distribution. In order to understand the variation 
in confusion at different limiting magnitudes $m_0$, fits were performed on sources 
brighter than respectively 18th, 19th and 20th magnitudes in $r$, in addition 
to fits across all sources.

In a perfect survey capable of deblending overlapping sources, all fields would
exhibit zero confusion, with their curves jumping from zero to 100\% of all
fields at a ratio of 1.0. In reality some sources will always be lost to
confusion, with the impact increasing -- the higher the density of objects. Fainter 
objects suffer more from confusion losses as their PSFs are more likely to be 
lost in the vicinity of brighter sources, in addition to their intrinsically higher 
densities (see Fig.~\ref{fig:magnitude_turnovers} for examples of the magnitude distributions 
of IPHAS fields). Fig.~\ref{fig:confusion_completeness} demonstrates this behaviour as the 
incompleteness takes hold for a greater number of fields as the cut-off magnitude $m_0$ 
is increased.

For a fraction of fields at all values of $m_0$, the best fit to their nearest
neighbour distributions indicates a lower predicted number of sources than
is actually observed. Indeed the median completeness at brighter $m_0$ shows signs of
converging to 0.98--0.99. However, we notice that the maximum ratio returned is an unphysical 
$\sim$1.1, for fields which lie in regions of lowest density in the map of the Plane. We take 
this as an indication that the fitting to the n$(\theta)$ distribution tails has an 
associated uncertainty of $\sim$10\% .  So large an 
uncertainty indicates that this method for correcting source counts is not itself sufficiently 
exact to adopt it for application across the survey.  
A significant contributing factor to this uncertainty is the adoption of a single 
nearest neighbour distribution for each IPHAS field. Marked variations or gradients in stellar 
density (due particularly to changing extinction) on scales of several arcminutes within fields 
will introduce error into the corrected count inferred from the tail of the nearest-neighbour 
distribution -- this is behind the unphysical under-prediction noted above.  Cutting fields up 
into smaller subregions would bear down on this source of error, at the price of reducing the
numbers of objects contributing to each distribution.  The more fundamental limitation on the utility
of this technique is that it does not easily provide completeness as a function of stellar apparent 
magnitude.

Nevertheless, Fig.~\ref{fig:confusion_completeness} can be regarded as a first demonstration that, 
as a general rule, at $r < 19$ the impact of confusion in IPHAS is small while, as might be 
expected, it is close to ubiquitous at $r > 20$.  Source counts in the $i$ band are 
higher thanks to the lower extinction -- the same plot for this band shows that 
confusion becomes a minor consideration once $i < 18$. 

\subsection{Other approximate measures of completeness}
\label{subsec:turnovers} 

A formula for correcting for the number of sources lost to 
confusion, that is based upon Eqn. \ref{eq:theoretical_neighbours}, may be written
\begin{equation}
\rho = - \rho^{\prime}\frac{\log\left(1-4\rho^{\prime}\pi\theta_{FWHM}^2\right)}{4\pi\theta_{FWHM}^2} 
\label{eq:irwin_confusion}
\end{equation}
\noindent where $\rho$ is the actual source density, $\rho^{\prime}$ is the observed density, and $\theta_{FWHM}$ is the seeing in which the field was observed. The expression was derived by \cite{Irwin1984}, and it was used to correct the observed n$(\theta)$ distributions directly in \citet{Gonzalez-Solares2008}. Clearly fields with poorer seeing and higher densities will be subject to larger corrections; \citet{Gonzalez-Solares2008} reported that the IPHAS Initial Data Release fields suffering from greatest confusion were missing 41\% of their sources.

\begin{figure}
\begin{center}
\includegraphics[width=1\linewidth]{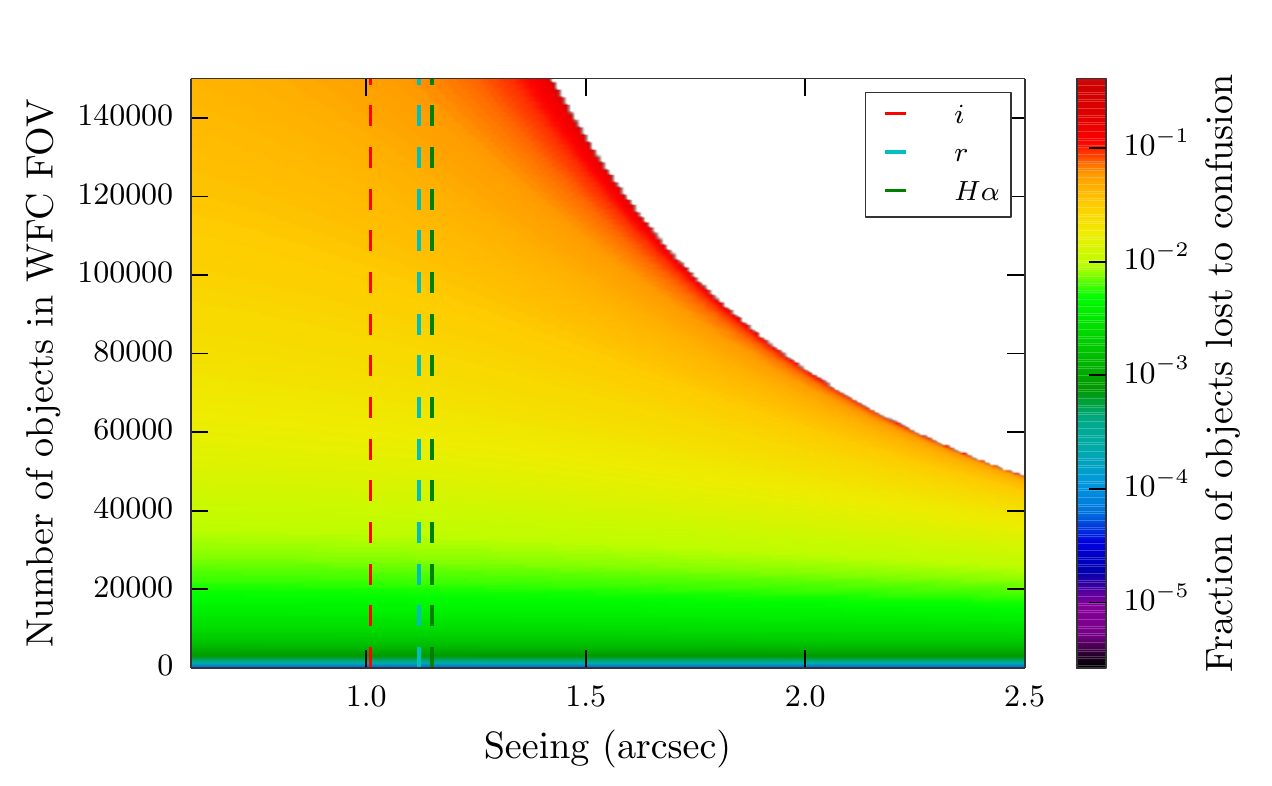} 
\caption{Correction term in Eqn. \ref{eq:irwin_confusion} for the range of $n_{src} = \rho^{\prime} \times \Omega_{WFC}$ and $\theta_{FWHM}$. Median seeings for each filter are marked by vertical lines. The white region denotes the domain where Equation \ref{eq:neighbour_breakdown} is true and Equation \ref{eq:irwin_confusion} breaks down. Note that vertical scale converts to the source count per square degree on dividing by the WFC footprint of 0.29 square degrees.}
\label{fig:neighbour_breakdown}
\end{center}
\end{figure}

However Eqn. \ref{eq:irwin_confusion} does not hold for for the entire
[$\rho^{\prime}$, $\theta_{FWHM}$] space covered by all IPHAS DR2 fields - it
breaks down in dense fields, where the seeing is appreciably worse than the median. Fig.
\ref{fig:neighbour_breakdown} shows the variation of the correction term for the range of parameters relevant
to DR2. The white region shows the parameter space in which Eqn. \ref{eq:irwin_confusion} is not 
applicable, where
\begin{equation}
4\pi\rho\theta^{2} > 1
\label{eq:neighbour_breakdown}
\end{equation}
Although this domain is only strictly entered by fewer than 20 fields, the performance of this approximate 
formalism -- well-suited to low density halo fields -- degrades as the domain is approached at the high densities
encountered away from the mid-Plane inside the Solar Circle. For this reason, its deployment across the survey is 
not the preferred option.

An approach used in several previous large-area surveys \citep{Ruphy1997, Cambresy2002, Lucas2008} 
is to estimate the completeness limit by taking the magnitude distribution of sources and
identifying at which magnitude the distribution begins to drop off. As part of a study of stellar 
populations in the Galactic Plane nearly two decades ago, \citet{Ruphy1997} noted that evaluating 
completeness based on recovery from simulated images is quantitatively the superior option, but for 
simplicity kept to the alternative of a completeness limit based on star-count histograms 
of uncrowded fields.   This approach may serve well in studies of limited sky regions, or if 
sources fainter than the deduced completeness limit can be set aside without causing problems. In 
this same study, confusion due to crowding was recognized as the dominant source of incompleteness, 
whilst ``slight variations due to the observing conditions" were judged to be relatively unimportant. 
This last presumption would be unsafe here.

\citet{Cambresy2002} estimated their completeness from the turnover of 2MASS
magnitude distributions, and reason that their density maps would show imprints
of individual observations had they overestimated their limiting magnitudes.
This would certainly be the case for IPHAS - without any magnitude cut in
place, the field-to-field variation of densities is extremely obvious.  In applying a 
similar treatment of incompleteness, \citet{Lucas2008} quote the 90\% completeness 
limit of the UKIDSS Galactic Plane Survey, noing that in uncrowded fields the modal
depths vary by 0.25 mag due to observing conditions. This method still depends
heavily on the turnover in the magnitude distribution, and relies on 
``visually extrapolating" the histogram.

A single completeness statistic as provided by \citet{Lucas2008} could be used
to apply a uniform completeness correction, although this assumes that all
fields in a survey suffer from the same pattern of incompleteness. Even if this
were indeed the case, only the count of sources brighter than the given
completeness limit could be corrected; it is not possible with this information
to correct the count going deeper.

\begin{figure*}
\begin{center}
\includegraphics[width=1\textwidth]{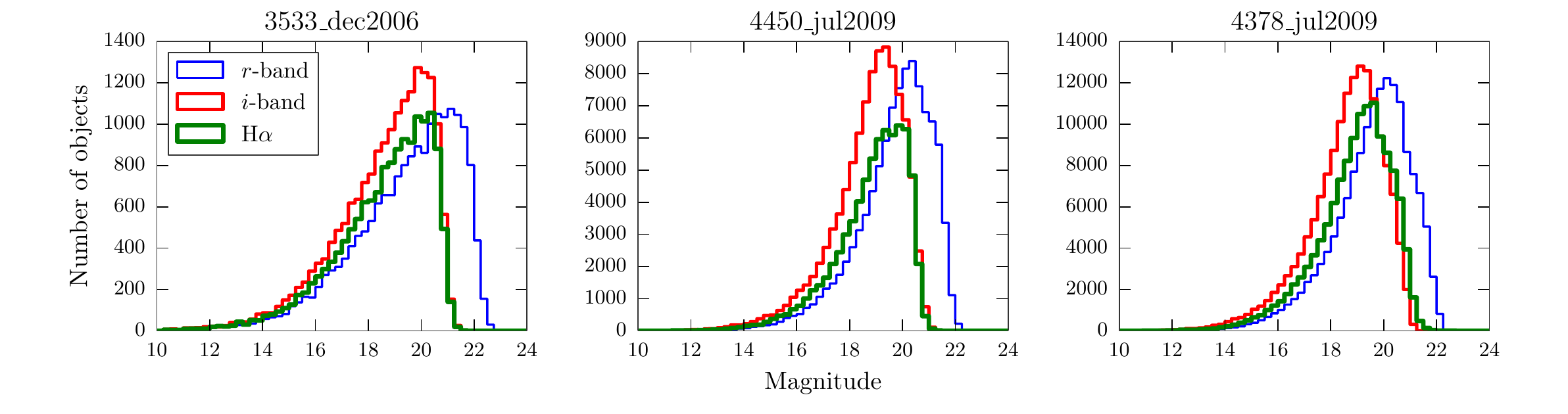} 
\caption{Magnitude distributions for three IPHAS fields of 
increasing density. From left to right, fields have $r$-band source 
counts of 18,113, 99,061 and 153,430. A requirement that sources counted here 
have $nBands>1$ places the greatest constraint on $i$-band counts, 
as redder sources are picked up in the $i$-band only.}
\label{fig:magnitude_turnovers}
\end{center}
\end{figure*}

Fig.~\ref{fig:magnitude_turnovers} shows the magnitude distributions for three IPHAS fields. 
A visual inspection would suggest that incompleteness sets in at $\sim$20th mag or fainter in
$r$ and at $\sim$19th in $i$ and $H\alpha$. Attempts to automate the determination of turnover 
magnitude for all fields are hampered by the facts that the mode of these distributions are 
dependent on the chosen binning, and that the magnitude distributions of many fields plateau 
before dropping off.  But it is already clear from the previous section and 
Fig.~\ref{fig:confusion_completeness} that the full picture is more complicated, with some small 
residual source loss still affecting magnitudes as bright as 18 to 19. In short, there is good reason 
to address correction rigorously on a field by field basis.

\subsection{Artificial source injection}  
A thorough treatment of incompleteness in any survey requires measuring its
sensitivity to sources over the entire magnitude range of interest - this is best achieved 
by simulating observations and then processing these simulated frames in the same way as 
the original data. This method (in addition to the magnitude distribution method 
discussed in Section~\ref{subsec:turnovers}) was used and described by \citet{Harvey2006} as part 
of a study of interstellar clouds observed by Spitzer.

Such an approach is much more powerful, as statistics can be returned on
magnitude bins, rather than on the entire magnitude distribution down to a
specified faint limit. It entails the simulation of sources of all
magnitudes, resembling the real data as closely as possible, whilst avoiding
undue cost in terms of both computing power or time.

For the purposes of correcting the density map, the completeness of each field
was assessed using the images and catalogues of CCD 4 only. Using only one of
the WFC CCDs per field cuts the processing time by a factor of four, bringing
the total time necessary to compute the completeness of the entire survey down
to a little over one week (using a high-performance multi-node cluster). Due to the 
fact that DR2 preferentially selects sources from around field centres (closer to the 
OA), CCD 4 -- the most central of the mosaic -- was selected as the chip 
to represent each field (see Section~\ref{subsec:completeness_fractions} for further 
discussion).

The main steps of the completeness calculation were as follows. First the
typical properties of an image were determined, and used to generate synthetic
sources that closely resemble genuine stellar sources (Section \ref{subsec:artificial_sources}). 
The parameters of these synthetic sources were recorded and then catalogues were generated, via 
the pipeline software, from the new frames with the added synthetic objects. The tables of 
synthetic source parameters were cross-matched against the newly generated catalogues, and the 
rate of recovery was measured (Section~\ref{subsec:recovery}). The fraction recovered at 
different magnitudes will allow a completeness curve to be built up for each field, resulting
in tailored corrections that can then be applied across the entire survey.

\subsection{Simulating stellar sources}
\label{subsec:artificial_sources}
The complexity of simulating stellar sources depends on the accuracy with which
the real sources are to be recreated. The simplest approximation of stellar images
to two-dimensional Gaussians, can be seen to be overly simplistic from a cursory 
glance at the reduced IPHAS images. Almost all IPHAS frames exhibit some degree of 
ellipticity (up to $e=0.2$ (see \citet{Barentsen2014}), requiring non-symmetric artificial 
sources. Elliptical 2D Gaussians were therefore chosen as sufficiently realistic profiles 
for artificial stellar sources (see Section~\ref{subsec:recovery} for further discussion).

The parameters returned by the aperture photometry performed on the DR2 images indicate the following 
should be included to accurately recreate a stellar source:
\begin{itemize}
\item Flux
\item FWHM
\item Ellipticity
\item Position angle
\item Coordinates (or pixel position on detector).
\end{itemize}
\noindent The choice of each is described below.

\subsubsection{Flux}
\label{subsubsec:flux}
The relation between total flux and magnitude was determined per field by fitting a power law to the photometry returned 
by \textsc{imcore}.

In order to insert a source of a given magnitude $m_0$, it is necessary to understand what flux in counts per unit time
that source should have. The magnitude of an object is determined by
\begin{equation}
m = ZP - 2.5 \times \log_{10} {\dfrac{c}{t}} - APCOR - PERCORR
\label{eq:aperture_photometry}
\end{equation}
\noindent where $ZP$ is the photometric zeropoint of the image as determined via the DR2 uniform calibration \citep{Barentsen2014}, $c$ is the measured counts within the defined aperture, $t$ is exposure time in seconds, 
while $APCOR$ and $PERCORR$ are small correction terms. $APCOR$ is an aperture correction term calculated by 
the pipeline software, \textsc{imcore}, which uses the curve-of-growth of stellar sources to determine the 
correction required to transform the chosen aperture measurement to total flux \citep{Irwin2001}. 
$PERCORR$ is a sky calibration correction, obtained by comparing dark sky regions with the median across each 
CCD. $ZP$, $APCOR$ and $PERCORR$ are all taken from the pre-existing catalogue headers.

\subsubsection{PSF shape parameters: Full-width at half-maximum and ellipticity}
\label{subsubsec:parameter_fwhm}

Initially the parameters of genuine sources in IPHAS frames were determined by
fitting elliptical two-dimensional Gaussian profiles to each source detected by
\textsc{imcore}, building up lists of best fit parameters for the purpose of
generating realistic artificial sources.  Parameters of sources were excluded from 
these lists if they exhibited one of the following behaviours: 
\begin{itemize}\item The source position returned is
further than 5 pixels from the position reported by \textsc{imcore} - these
fits are likely to have been disturbed by nearby sources \item The
peak value of the best-fit Gaussian is $>55000$ counts - this is the regime
for the WFC where bright sources saturate, distorting the PSF. \end{itemize}

The best fit full-width half maximum (FWHM) values returned for the semi-major and 
semi-minor axes of the sources were gathered for each of the fields entering this 
exploration, and Gaussians were then fit to the distribution to recover the distribution
mean and $\sigma$. The FWHM value representative of each field returned by \textsc{imcore}
fell within 1$\sigma$ of the mean, confirming that the pipeline-specified 
value can be re-used safely for articial source generation.

Similarly, the measure of field-wide ellipticity returned by \textsc{imcore} was found to 
agree with the typical ellipticities obtained in the process of fitting Gaussians to each object across the field using the relation
\begin{equation}
e=1-\frac{\mathrm{FWHM}_{min}}{\mathrm{FWHM}_{maj}}
\label{eq:ellipticity}
\end{equation}
\noindent (where FWHM$_{min}$ and FWHM$_{maj}$ are the full-width half-maxima along sources' semi-minor and semi-major 
axes respectively). As with FWHM, the \textsc{imcore} ellipticity value per field was used in artificial source generation.

\subsubsection{Position angle}
\label{subsubsec:position_angle}

The typical distribution of position angle (PA) within a field is quite broad and non-Gaussian in
appearance.  In practice, there is no reason to expect the choice of PA to significantly influence 
the ability to recover an artificial source -- hence it is set at zero for fields where no obvious peak 
in the PA distribution was observed, otherwise artificial source PAs were drawn from the best fit 
to the peak of the PA distribution.

\subsubsection{Position}
\label{subsubsec:position}
Object $x$ and $y$ pixel positions were drawn randomly from uniform
distributions across the 2048$\times$4096 pixels of the WFC CCDs.

The completeness calculated for the unvignetted CCD 4 of each field was taken to represent 
the completeness over all four CCDs.  This allowed the border zone -- which can complicate the 
appraisal of completeness -- to be easily defined: a rectilinear zone 10 arcsec (or 30 pixels) wide
inside the CCD edges was imposed such that no artificial source could be placed within it.  This 
made sure that no injected artificial source would lose flux across the CCD edge.

\subsection{Practical implementation of artificial source injection}

Ideally measuring the completeness using artificial sources would be done by adding a single source 
at a time, verifying whether or not it is detected. This process would be repeated as many times as required 
to obtain decent statistics, for the entire magnitude range of interest. This would be the route taken in the 
absence of any limitations on either time or computing resources.   The limiting factor here
is the computing time taken for the basic step of catalogue generation (20 sec).  To generate the 
number of sources detailed in Table \ref{table:number_artificial} it would take $\sim$ 65 hours to process 
just one field.

To reduce the cost, multiple sources must be inserted into each image simultaneously. In practice, it was 
necessary to also ensure that too many sources were not inserted at any one time, since too high a number would modify the intrinsic properties of the image. For example, probing the completeness of an originally sparse image with a high number of artificial sources inserted would not return statistics useful for understanding the 
original image. The value of $\frac{\delta n}{n_{image}}$ needed to be kept sufficiently low, where 
$\delta n$ is the number of artificial sources added to the image, and $n_{image}$ is the number of sources 
present originally. The quality control information available for DR2 fields report that the most sparsely 
populated fields contain more than 1000 stars. This is not an extremely constraining limit; a maximum of up to 
50 stars was chosen as a value that would keep $\frac{\delta n}{n_{image}}$ below 0.05 for all fields.   
Allowing up to 50 sources per artificial source injection increases processing efficiency, but offers too few 
to sample the pre-existing magnitude distribution of the sources in each image faithfully.  Hence, the 
approach adopted was to split the magnitude range of interest into bins $0.25$ mag wide, inserting sources from 
only one bin at a time in each catalogue regeneration.

\begin{table}
\begin{center}
\begin{tabular}{|c|c|c|c|c|}
\hline
\multicolumn{2}{|c|}{Magnitude bin} & & & \\ \hline
Start & End & $N$ & $M$ & $No.\ sources$ \\ \hline
11.0 & 11.25 & 10 & 10 & 100 \\
11.25 & 11.5 & 10 & 10 & 100 \\[-1.5mm]
\vdots & \vdots & \vdots & \vdots & \vdots \\
13.5 & 13.75 & 10 & 10 & 100 \\
13.75 & 13.0 & 10 & 10 & 100 \\ 
14.0 & 14.25 & 20 & 10 & 200 \\[-1.5mm]
\vdots & \vdots & \vdots & \vdots & \vdots \\
15.25 & 15.5 & 20 & 10 & 200 \\
15.5 & 15.75 & 30 & 10 & 300 \\[-1.5mm]
\vdots & \vdots & \vdots & \vdots & \vdots \\
16.75 & 17.0 & 30 & 10 & 300 \\
17.0 & 17.25 & 40 & 10 & 400 \\[-1.5mm]
\vdots & \vdots & \vdots & \vdots & \vdots \\
17.75 & 18.0 & 40 & 10 & 400 \\
18.0 & 18.25 & 50 & 10 & 500 \\
18.25 & 18.5 & 50 & 10 & 500 \\
18.5 & 18.75 & 50 & 15 & 750 \\
18.75 & 19.0 & 50 & 15 & 750 \\
19.0 & 19.25 & 50 & 20 & 1000 \\[-1.5mm]
\vdots & \vdots & \vdots & \vdots & \vdots \\
19.75 & 20.0 & 50 & 20 & 1000 \\\hline
\multicolumn{3}{|r|}{Total:} & 410 & 12300 \\\hline
\end{tabular}
\caption{Number of artificial sources added per $i$-band magnitude bin. $No.\ sources$ denotes total number that will be generated over $M$ runs. $N$ is the number of sources that will be added to each image, which will be repeated $M$ times. A total of 12300 are added per field, across 410 images containing artificial sources. For the $r$-band, the bins were shifted 1 mag fainter.}
\label{table:number_artificial}
\end{center}
\end{table}

Table \ref{table:number_artificial} details $N$, the number of sources inserted per image, and $M$, the number 
of artificial images generated for reprocessing and recovery for each magnitude bin. $N$ rises  
with increasing magnitude and is capped at 50 at the faint end for the reasons discussed above. To accumulate a 
larger number of sources injected in the faintest bins, and minimize noise where the most significant corrections 
arise, $M$ increases from 10 to 15 at $m_0 = 19.5$, and then to 20 at $m_0 = 20.0$.  This setup requires around 
2.2 hours ($\sim$20 s $\times$ 410 \textsc{imcore} runs). For the 14,115 fields that make up DR2, this results in 
a total computing time of $\sim$31000 hours. Utilizing 128 CPUs simultaneously, as was done here, brings the total time to estimate completeness for the entire survey in a single band to $\sim$10 days.

\subsection{Recovery of simulated sources}
\label{subsec:recovery}

Cross-matching the recovered sources to the added artificial sources required the imposition of matching thresholds 
in both position and brightness. While, ideally, a recovered source should lie centred on the exact $x$,$y$ pixel 
position where it was added, there is no guarantee that \textsc{imcore} will report unchanged centroid coordinates. 
Background variations in the original image, pixel binning and blending may cause the reported centroid to shift 
by a small amount: hence, a generous upper limit of 5 pixels (compared to a typical shift of less than one pixel) was set on this displacement. 

Simply cross-referencing the recorded position of an artificial source with the regenerated catalogue is not 
sufficient to confirm successful recovery. A constraint on the difference between the recovered magnitude, $m_{recovered}$ 
of the source and its insertion value $m_{synthetic}$ is essential, in order to reduce spurious crossmatches between added 
sources and pre-existing nearby sources (typically of different brightness).  Choosing a bound on this offset, 
$\Delta_m = m_{synthetic} - m_{recovered}$, was more involved than for pixel position.  Previous attempts to use artificial 
photometry to understand incompleteness have faced similar issues.  For example, \citet{Mateo1986}, as part of 
a study of a globular cluster in the Large Magellanic Cloud, added artificial stars to images using 
\textsc{DAOPHOT}: they considered a source recovered if it was returned at the same coordinates at a 
magnitude within 0.5 of the inserted value.  \citet{Harvey2006} used this approach in a study of stellar 
sources in the Serpens molecular cloud, and plotted the modulus of the average difference for artificial 
sources injected into their data, finding a range from 0.1 mag at $\sim$10-13 mag to more than 0.5 mag at 
$\sim$15 mag.

\begin{figure}
\begin{center}
\includegraphics[width=1\linewidth]{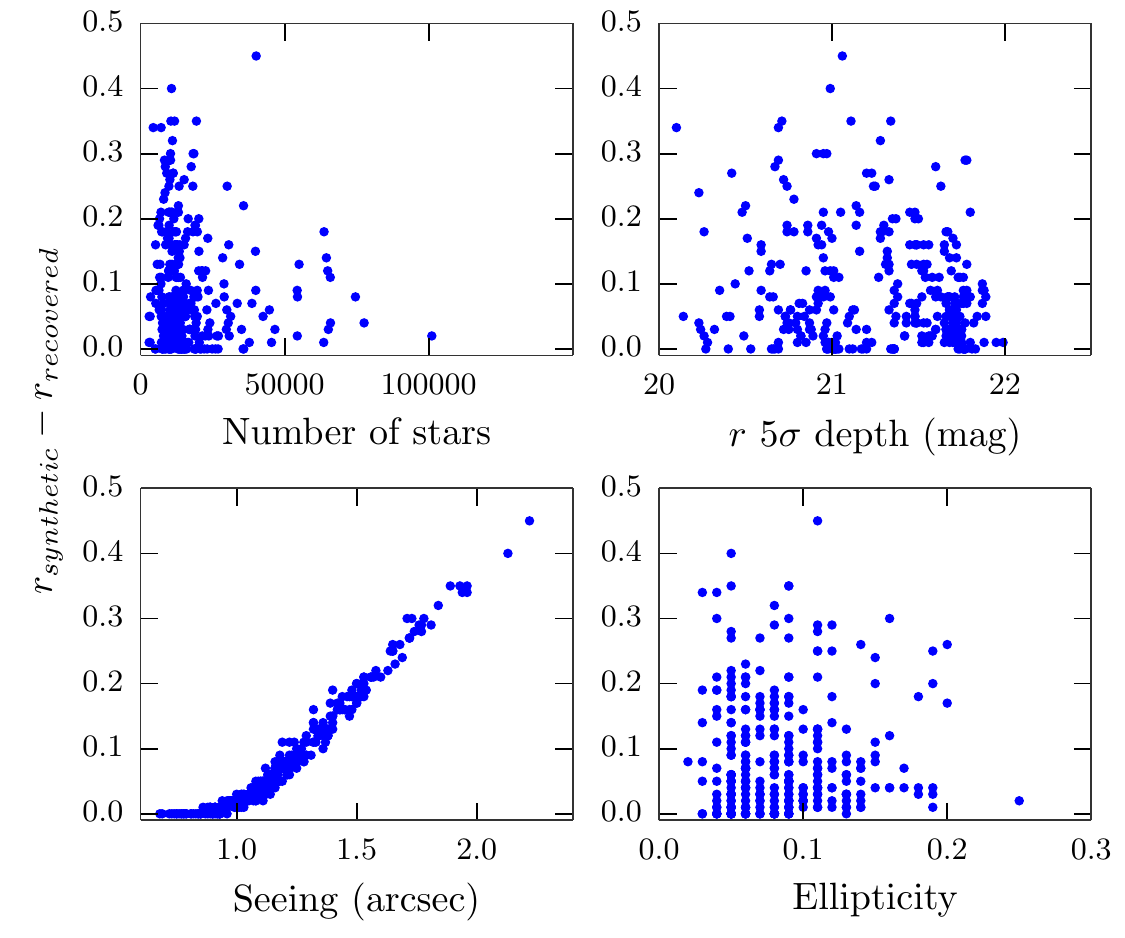} 
\caption{Modal difference between artificial inserted and
recovered magnitude per field, plotted against field
parameters.}
\label{fig:magdiff_parameters}
\end{center}
\end{figure}

For each field, the distribution of the modulus of $\Delta_m$ was binned up and the 
modal value determined. For a number of fields, the modal value reached $\approx$0.4
mag - an effect that would require a $\Delta_m$ tolerance of $>$0.5 mag. Such
a large tolerance would likely result in many spurious cross-matches and hence
an unduly optimistic estimate of completeness fractions, especially at fainter
magnitudes.  To better understand this, the modal $\Delta_m$ for each field
was plotted against a number of field parameters. The resulting plots, seen in
Fig.~\ref{fig:magdiff_parameters}, clearly show this quantity correlates most strongly 
with the PSF FWHM (the `seeing').  Indeed, the relationship underlying the tight trend 
is that the returned magnitudes from a regenerated catalogue are offset by a systematic 
amount rather than spread widely around the mode - a pattern that persists, up to the highest 
PSF FWHM encountered in the survey. This finding has motivated the application of a 
seeing-dependent uniform shift to all recovered magnitudes from an image, applied to bring 
the median $\Delta_m$ of a field back to zero. The origin of this offset is likely to be  
the difference between the stellar profiles of the artificial and the real sources. 
In this sense the empirically determined offset and its correction are the trade for having 
used Gaussian profiles in place of more complex Moffat profiles.

\begin{figure}
\begin{center}
\includegraphics[width=1\linewidth]{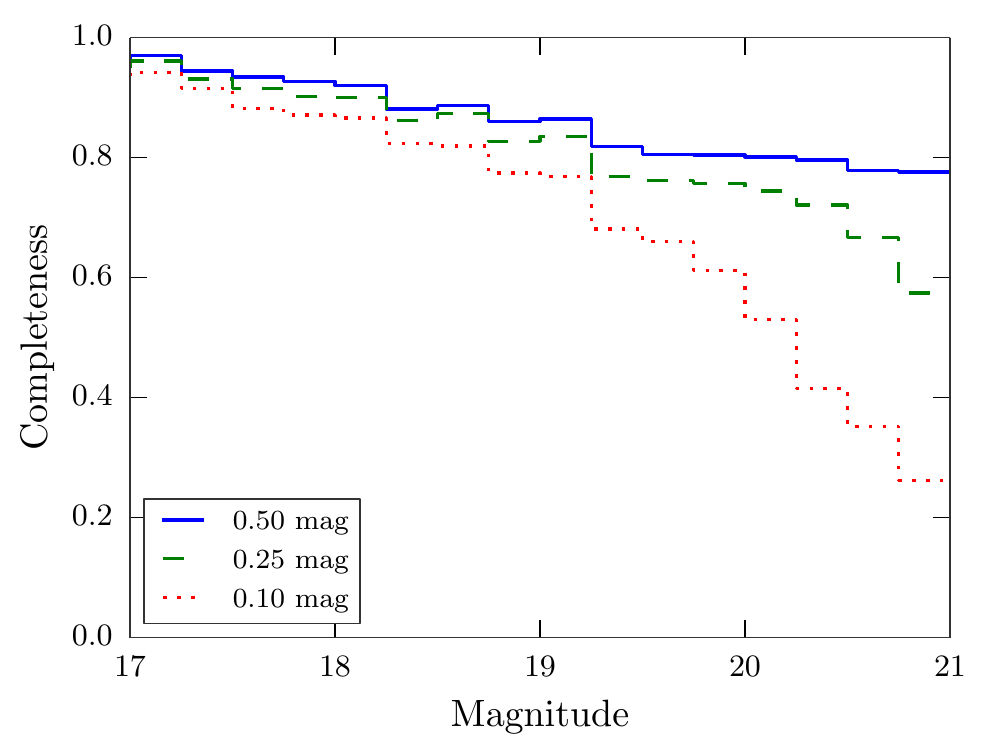} 
\caption{An example of $r$-band completeness fractions for a single field exposure 
obtained using offset tolerances between inserted and recovered magnitudes of 0.5 (blue solid), 
0.25 (green dashed), and 0.1 mag (red dotted).  The middle value of 0.25 is the adopted tolerance.}
\label{fig:magnitude_match}
\end{center}
\end{figure}

After taking out the systematic shifts to $\Delta_m$ values, a bound on the acceptable
corrected difference $\Delta_m^{cor}$ still needs to be set. As illustrated in 
Fig.~\ref{fig:magnitude_match}, the completeness curves
remain relatively unchanged at $r\lesssim19$ as the $\Delta_m^{cor}$ limit is
varied between 0.5 and 0.1 mag. This limit was set at 0.25 mag, as a compromise
between avoiding acceptance of a large number of spurious detections at the faint
end of the magnitude distribution, and ensuring that few objects are missed
due to the occasional larger $\Delta_m^{cor}$ value.

\subsection{Completeness fractions}
\label{subsec:completeness_fractions}

\begin{figure*}
\begin{center}
\includegraphics[width=1\textwidth]{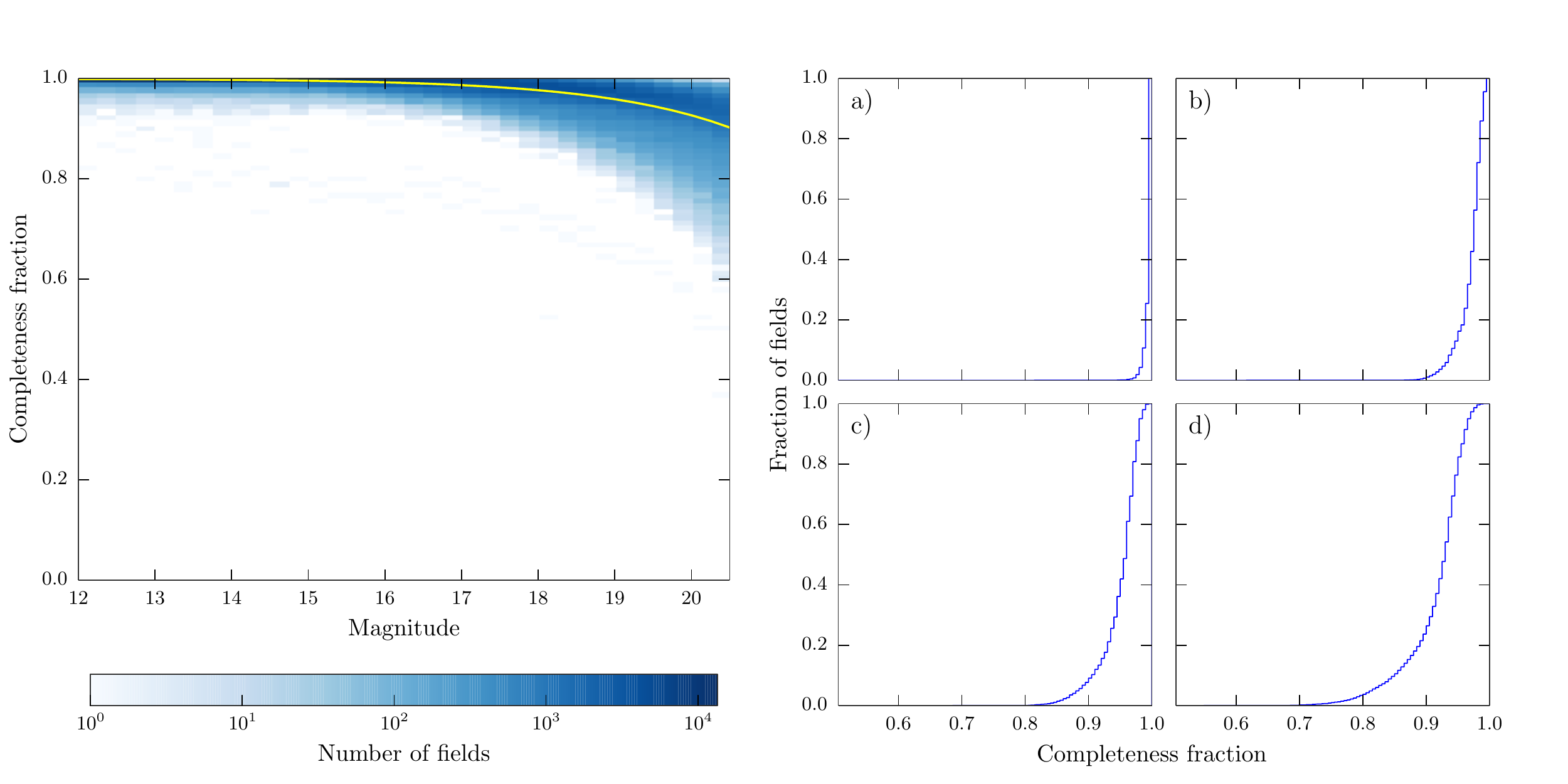} 
\includegraphics[width=1\textwidth]{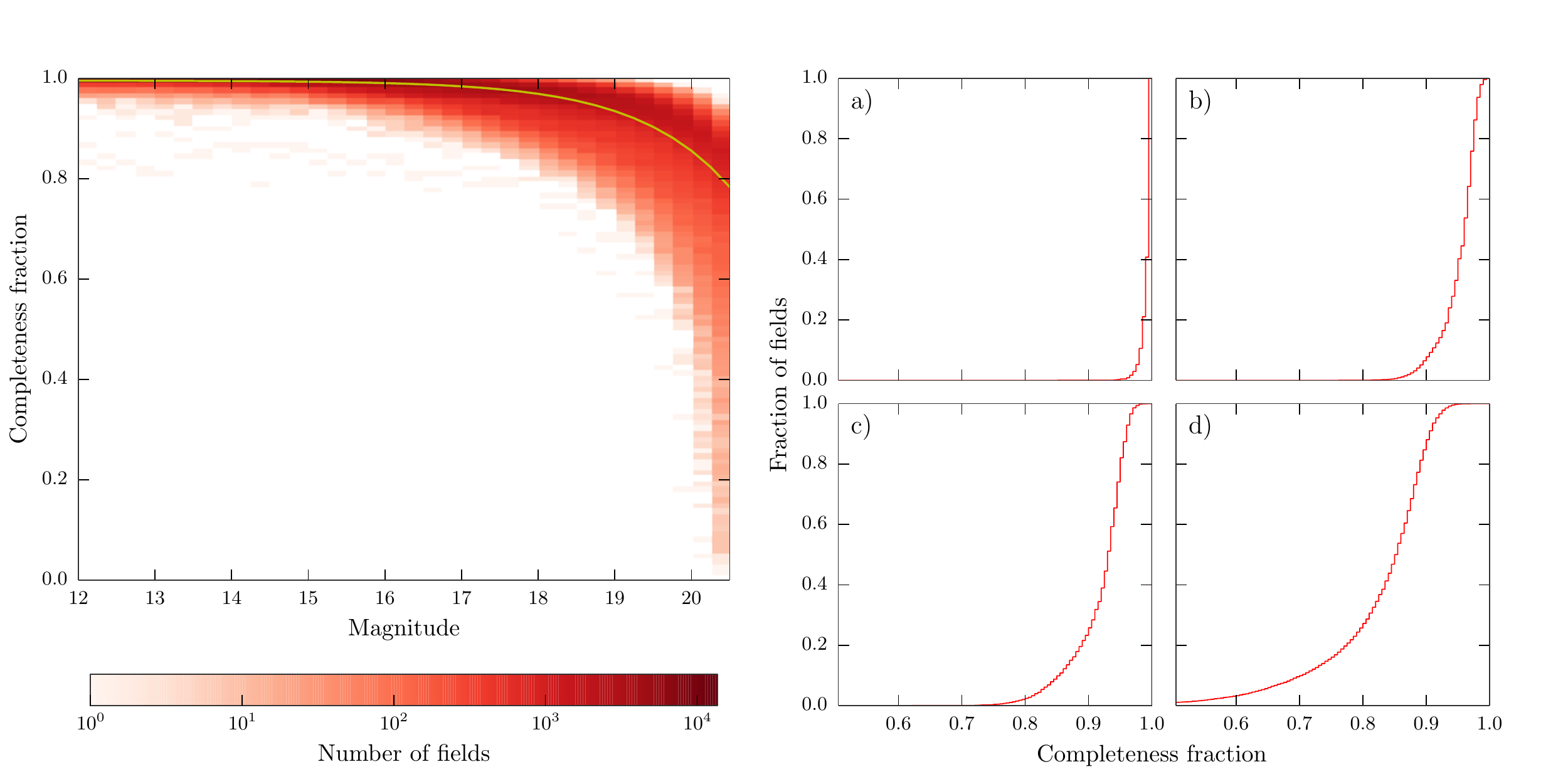} 
\caption{\textbf{Left:} A density plot incorporating every completeness curve in DR2, for magnitude bins of width 0.25 in between 12th and 21st mag in $r$ (\textbf{upper}) and $i$ (\textbf{lower}). The yellow line superposed in each panel is the fit curve obtained using eqn.~\ref{eq:completeness_curve} with the median value for each fit parameter. \textbf{Right:} Cumulative histograms of field completeness for magnitude bins \textbf{a)} 15.0-15.25 \textbf{b)} 18.0-18.25, \textbf{c)} 19.0-19.25, \textbf{d)} 20.0-20.25.}
\label{fig:completeness_curves}
\end{center}
\end{figure*}

Fig.~\ref{fig:completeness_curves} shows the completeness curves of all DR2 fields; the discrete quantisation of 
completeness is more prominent at the bright end due to fewer artificial sources being injected to assess the 
incompleteness of bright sources. The upper two cumulative histograms shown in the upper set of panels in Fig.~\ref{fig:completeness_curves} show that incompleteness remains low at bright magnitudes, with a median completeness of 99.5\% at 
$r$=15. At 18th magnitude the median completeness of IPHAS is 97.7\% , falling to 93.1\% at 20th magnitude.  The pattern 
in the $i$ band (lower panels) is similar, except that completeness declines more rapidly with increasing magnitude such that the data for $i$ at 19th magnitude more closely resembles $r$ at 20th than $r$ at 19th.

\begin{figure}
\begin{center}
\includegraphics[width=1\linewidth]{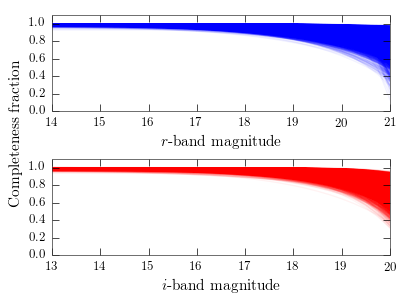} 
\caption{Best fits to completeness curves for both $r$ (\textbf{upper}) and $i$ (\textbf{lower}) catalogues.}
\label{fig:completeness_fits}
\end{center}
\end{figure}

In order to apply completeness corrections when generating the density map, each detected source needs to contribute to the map by a factor that takes into account the incompleteness of sources of similar magnitude from its field of origin. In order to characterize the incompleteness fraction of each field as a quantity varying continuously with magnitude ($r$ as written below), a function was fit to the measured completeness fractions of each field, of the form
\begin{equation}
C(r) = \alpha - \gamma \times \mathrm{e}^{\frac{r}{\beta}}
\label{eq:completeness_curve}
\end{equation}
\noindent where $\alpha$, $\beta$, and $\gamma$ are parameters allowed to vary to find the best fit for each field. These parameters were collected and formed a lookup table for use in correcting the density map. The curves of this form can be seen plotted for every field in both $r$ and $i$ in Fig.~\ref{fig:completeness_fits}. Median values for the coefficients are $\alpha=0.996, \beta=1.20, \gamma=8.08\times10^{-9}$ for the $i$-band, and $\gamma=0.999, \beta=1.73, \gamma=6.89\times10^{-7}$ for the $r$-band. The curves corresponding to these median values have been plotted in Fig.~\ref{fig:completeness_curves}.  A table containing sets of coefficients for all DR2 fields is available online\footnote{http://www.iphas.org/data/dmap}.

As mentioned in Section \ref{subsec:artificial_sources}, the completeness curves generated from CCD 4 of each field were used to represent those of the field as a whole. To ensure that this approach was sensible, a randomly-selected set of fields had completeness curves generated for all four CCDs in order to compare completeness curves and ensure that the variation between chips was acceptably low. For the magnitude range $12<r<21$, standard deviations in the completeness corrections were calculated. At $r=19$, the deviation in completeness corrections between CCDs reaches as high as 0.025, where a bright star appears in one or more CCDs of the field. But these cases are rare: the median $\sigma$ at $r=19$ 
is 0.005.

\section[]{The density maps}
\label{sec:dmap}

\subsection{Source counting and application of the incompleteness corrections}
\label{subsec:counting}
The IPHAS footprint was split into $1^{\prime}\times1^{\prime}$ cells, and for each cell, a table identifying 
the extent of every image contributing to DR2 (at the CCD level, i.e. 4 CCD footprints per WFC exposure) was 
queried to identify which intersect the cell (either completely or in part). In order to 
calculate the effective footprint of each CCD image, the pixel coordinates of its corners were determined, and a customized unusable border region was put in place.  This served mainly to exclude sources detected far from the WFC's OA where the counts are less reliable. 

For each cell, the relevant files identified in the coverage table were accessed, and all sources falling within the cell boundaries were selected. Sources meeting the following criteria were retained:
\begin{itemize}
\item Morphology classification -1, -2 or +1
\item Brighter than the adopted faint limit
\item Flag errBits $<$ 64 for band of interest
\end{itemize}
\noindent where the errBits criterion eliminates sources affected by one or more of: bad pixels within the PSF; a truncated PSF, vignetting. The exclusion of borders
already eliminates the majority of such cases, and so mainly leaves to this criterion the job of removing sources affected by issues such as bad columns and hot pixels.

For each selected source, a \textit{corrected contribution} to the number count of the cell was computed from the 
completeness curves generated in Section \ref{subsec:completeness_fractions}. The table containing all $\alpha$, $\beta$, 
and $\gamma$ values (Eqn. \ref{eq:completeness_curve}) was read in, so that the values relevant to the 
current CCD footprint can be identified.  Next, the magnitude of the source under consideration was used with these parameters to determine the source's corrected weight. The original ($\phi$) and corrected ($\Phi$) source counts were recorded, along with the fraction of the cell overlapped by the CCD footprint from which the source detections are drawn. Each estimate of total corrected cell occupation, $\Phi^{\prime}$ is given by 
\begin{equation}
\Phi^{\prime} = \Phi / f_c
\label{eq:corrected_density}
\end{equation}
\noindent where $f_c$ is the fraction of the cell covered.  A formal error estimate would be the Poisson noise 
of the observed area scaled to the cell:
\begin{equation}
\Phi^{\prime}_{err} = \sqrt{\Phi}/f_c \ \ .
\label{eq:corrected_poisson}
\end{equation} 
\noindent This is repeated for each CCD footprint overlapping the specified cell, and for all cells in the density map.  

At $1^{\prime}\times1^{\prime}$ resolution, the density map contains 6,537,051 cells overlapping with IPHAS 
photometry. Of these, 92.9\% overlap two CCD footprints, with 38.3\% overlapping three. The cells benefitting from repeated observations provide empirical estimates of the variance in source counts between observations, which in turn can be compared with the formal Poisson uncertainty (Eqn.~\ref{eq:corrected_poisson}), discussed further in Section~\ref{subsec:count_uncertainty}. In cases where multiple exposures and detections are available, the count from the CCD providing the best coverage fraction was adopted, with the observation made under the best conditions chosen from sets of observations providing equal coverage.

\begin{figure}
\begin{center}
\includegraphics[width=1\linewidth]{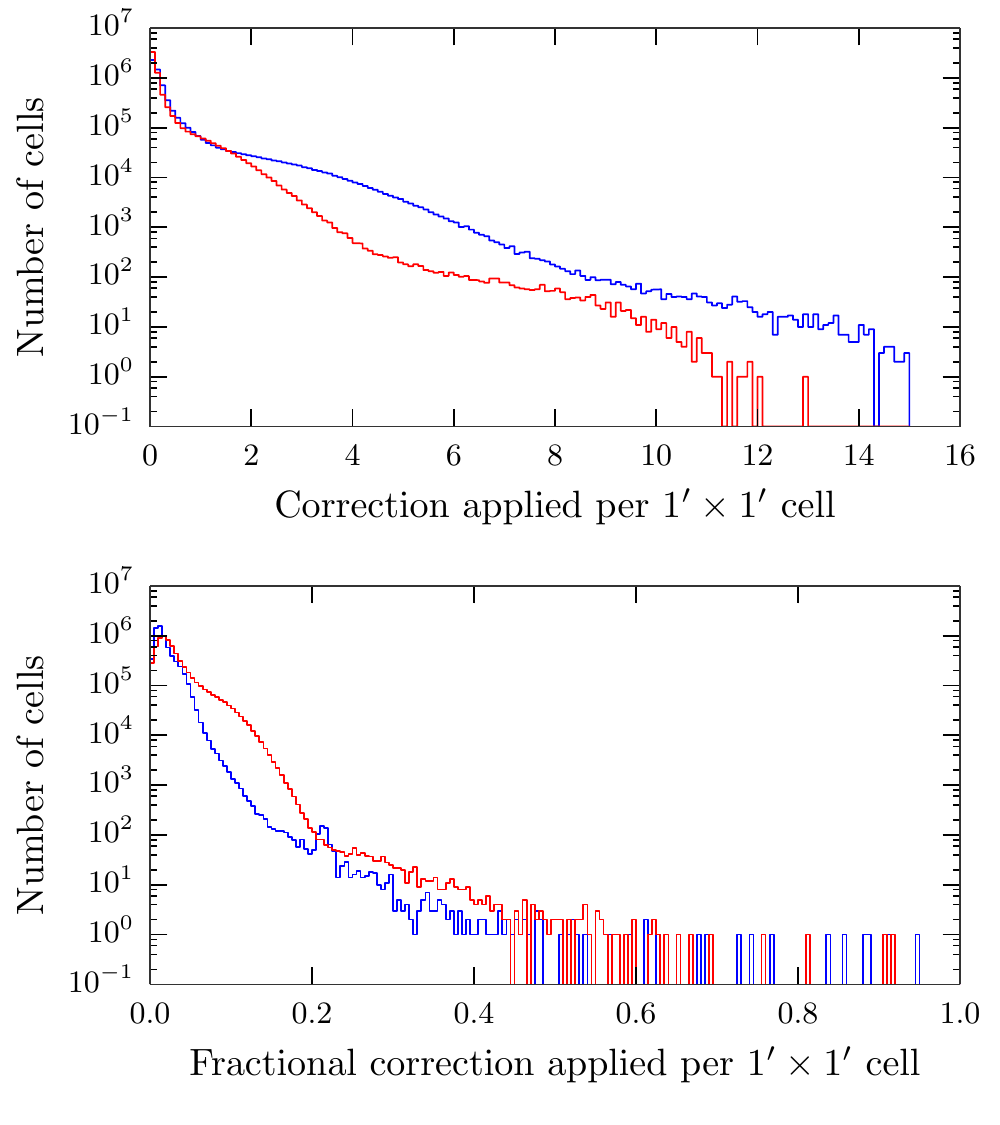} 
\caption{Histogram of the incompleteness corrections. \textbf{Upper}: expressed as the number of added sources. \textbf{Lower}: exposed as the fractional correction per cell applied to the $1^{\prime}\times1^{\prime}$ $r=19$ (blue) and $i=18$ (red) density maps.}
\label{fig:corrmap_hist}
\end{center}
\end{figure}

Fig.~\ref{fig:corrmap_hist} shows the histogram of corrections for the $r=19$ and $i=18$ density maps, confirming that most cells require an additive count correction that is well under 1.  

\subsection{Bright stars}
The fact that the $r$- and $i$-band maps are generated independently means that spurious sources 
detected in one filter only can be included in the map. The majority of such sources are noise and as such are eliminated based on their morphological classification. However in the regions surrounding bright stars 
($V\lesssim3$, particularly), the scattered light produced can lead to a large number of spurious detections that finish up classified as stellar or extended sources. Crossmatching between bands to eliminate such detections is 
not an option, as this would eliminate redder sources included in the $i$-band map.

Around fainter, but still bright, stars ($V\lesssim5$) the scattered light is not so severe as to increase the 
number of spurious sources; instead, a zone of missing sources is likely due to the bright PSF wings of these stars. This incompleteness is more localized than taken into account by the approach of Section~$\ref{sec:completeness}$.  To deal with these issues, cells lying within $5^{\prime}$ of stars brighter than $V=5$ appearing in the catalogue of \citet{Hoffleit1991} were simply excluded from the density map. The resultant cutouts around 102 stars has resulted in the discarding of 0.2\% of the density map cells.

\subsection{Uncertainties in the star counts}
\label{subsec:count_uncertainty}

\begin{figure}
\begin{center}
\includegraphics[width=1\linewidth]{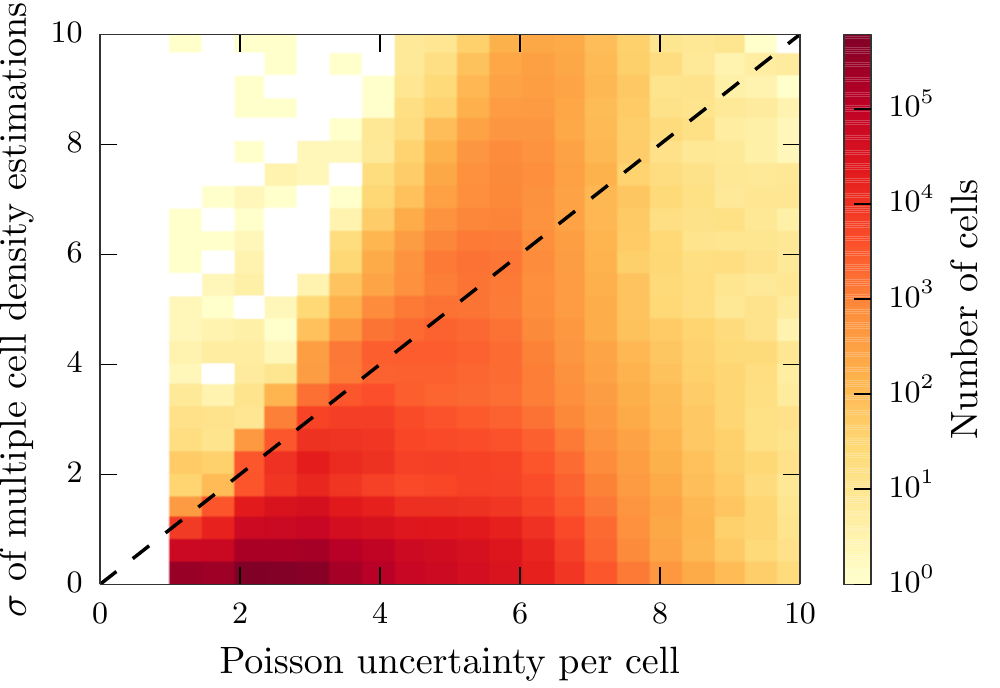} 
\caption{Density plot comparing the deviation between repeated source counts and poisson uncertainty for $\sim$4 million cells of the $1^{\prime}\times1^{\prime}$ corrected $i=18$ density map. These cells are those which are covered by multiple IPHAS observations, where the coverage fraction of all observations is greater than 0.5. The equivalent plot for the $r=19$ density map uncertainty appears indistinguishable from the $i=18$ - this is not unexpected; the typical $r-i$ value of ~1 suggests the two maps are capturing the same population.}
\label{fig:error_comparison}
\end{center}
\end{figure}

For sky cells captured in more than one exposure, the availability of repeated source counts provides a pessimist's measure of the impact of the observationally-driven error, existing in the density maps at the level of the $1^{\prime}\times 1^{\prime}$ cells (after correction for incompleteness). It is conservative in the sense that variance computed between counts from the different independent and independently-corrected observations pays no heed to the relative qualities of the compared observations.  It is significant then that this variance is nevertheless, in the vast majority of cases, distinctly lower than the scaled Poisson uncertainty (Eqn~\ref{eq:corrected_poisson}) on the adopted cell count - as shown in Fig.~\ref{fig:error_comparison}. More specifically, down to the $r = 19$ limit, for cells covered multiple times with coverage fraction greater than 0.5, the median deviation between cell source counts is 0.03 compared to a median Poisson uncertainty of 2.68.  Clearly simple Poisson statistics do not apply.  To this same limit the median source count is $\approx$6.2 (which scales up to $\approx$20000 sources per square degree). These numbers are similar for the $i=18$ limit, with a median deviation between repeated cell counts of 0.03, and a median Poisson uncertainty of 2.66. The ratio, 0.03/6.2, points to a typical error at maximum resolution of the map of $\lesssim$0.5\% due to the source count derived from IPHAS observations.  Rebinning the map to coarser resolution will bring this typical error down. 

Special measures are needed for deciding whether empty cells are reliable or not.  First, cells which are fully covered by IPHAS but contain no sources (i.e. genuinely empty cells down to the adopted faint limit) have an uncertainty placed on them equal to the contribution of a single source at the faint limit of the density map. Continuing the example of the $r$ map to 19th magnitude, the median source count of 6.2 in $1^{\prime}\times1^{\prime}$ density map cells indicates, on average, that 16\% of the area of any one cell of this size needs to be covered in order to provide a single source detection - if the fraction covered is less than this, a null source count should be regarded as unreliable. We use this reasoning to identify and weed out unreliable empty cells: specifically, on multiplying the expected uncertainty for an empty cell (typically $\approx1.2$ at $r=19$) by the minimum acceptable cell coverage fraction (rounding it to 20\%), a limiting scaled error of 6 counts is obtained for it. Any unoccupied cells with computed (Poisson) errors exceeding this level were discarded as being unreliable.  This has meant the retention of 22500 genuinely empty cells, and the rejection of 16352.  Genuinely empty cells are most commonly found in regions associated with high extinctions (e.g. towards the Aquila and Cygnus Rifts). The rejected cells mostly trace out the edges of fields, and are preferentially located at higher Galactic Longitudes, where the majority of gaps in IPHAS DR2 coverage fall.

The main causes of error in the derived stellar density in any one cell (with high coverage fraction) are the uncertainty in the completeness correction and the imprecision in the applied magnitude calibration.  The completeness correction errors can be kept as low as computing power permits -- here, the many hundreds of artificial sources in the fainter bins (see Table~\ref{table:number_artificial}) would imply errors in the completeness fraction estimates on the order of 3\% to 5\%.  How this feeds through to the corrected density maps will scale as the counts correction: e.g. a high-end 30\% counts correction to a cell would add under 2\% to the error budget (the fractional count corrections are mapped out in the Appendix).  Turning to the magnitude calibration, the external precision of IPHAS DR2 $r$ and $i$ is formally in the region of 0.03 magnitudes (see \citet{Barentsen2014}).  Viewing this conservatively by allowing that the magnitude cuts applied to the density maps may be wrong by as much as 0.05, an uncertainty in source count of $\sim$2\% would be deduced. These factors together imply a small total error budget in the region of 3\% that applies at any density map resolution and to both the $r$ and $i$ bands. This source of error, being greater than the source count errors discussed, is therefore the dominant source of uncertainty in the final density maps. Errors of no more than $\sim3-4\%$ should be borne in mind when studying source densities derived from the maps (discussed in Section \ref{sec:galactic_predictions}).

\subsection{The final stellar density maps}
\label{sec:the_maps}

\begin{figure}
\begin{center}
\includegraphics[width=0.95\linewidth]{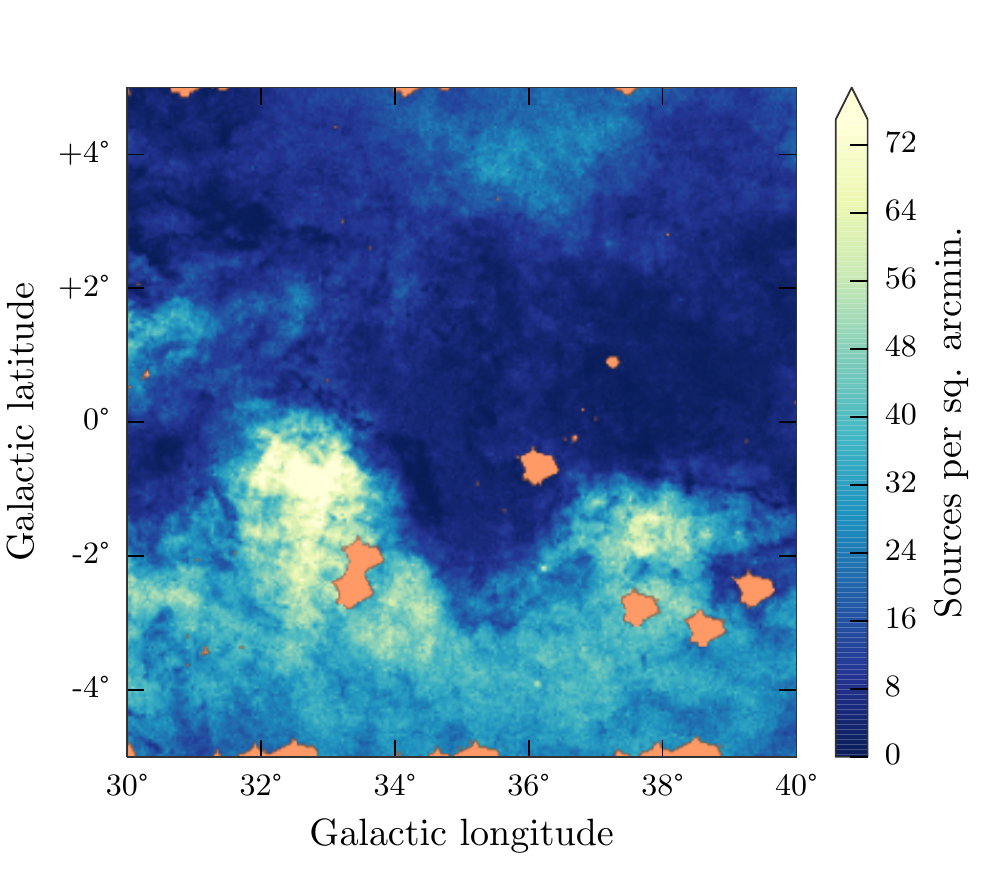} 
\includegraphics[width=0.95\linewidth]{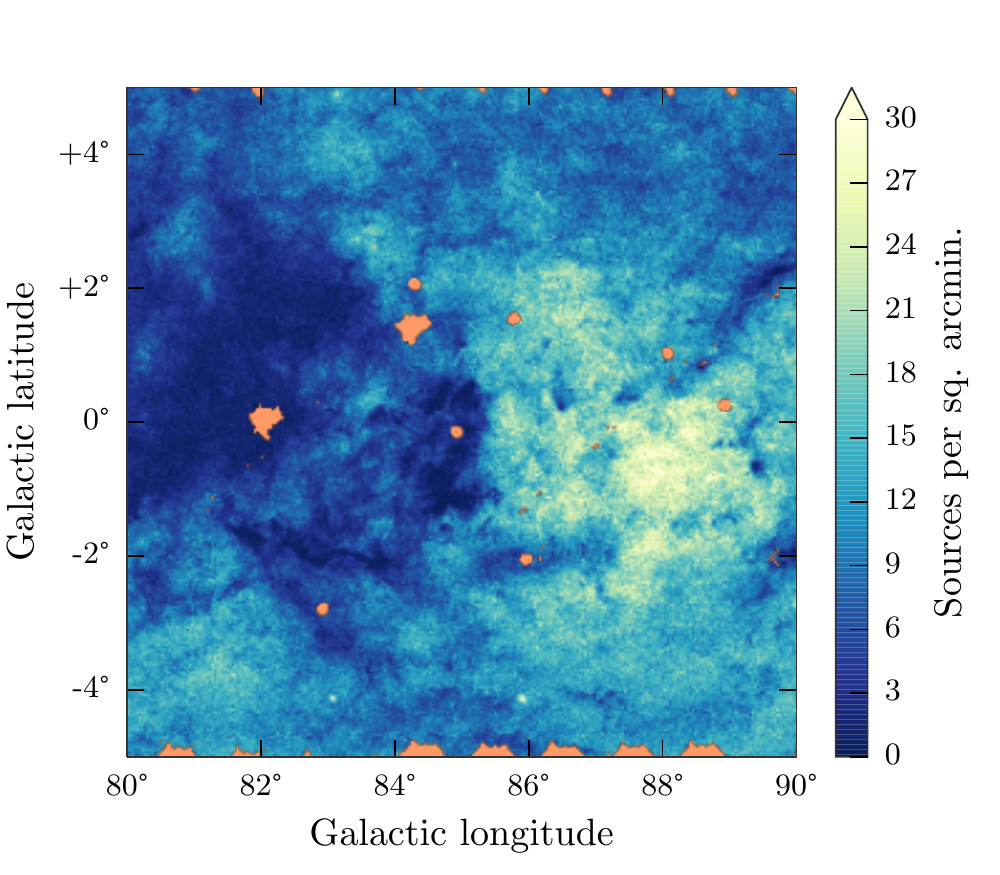} 
\includegraphics[width=0.95\linewidth]{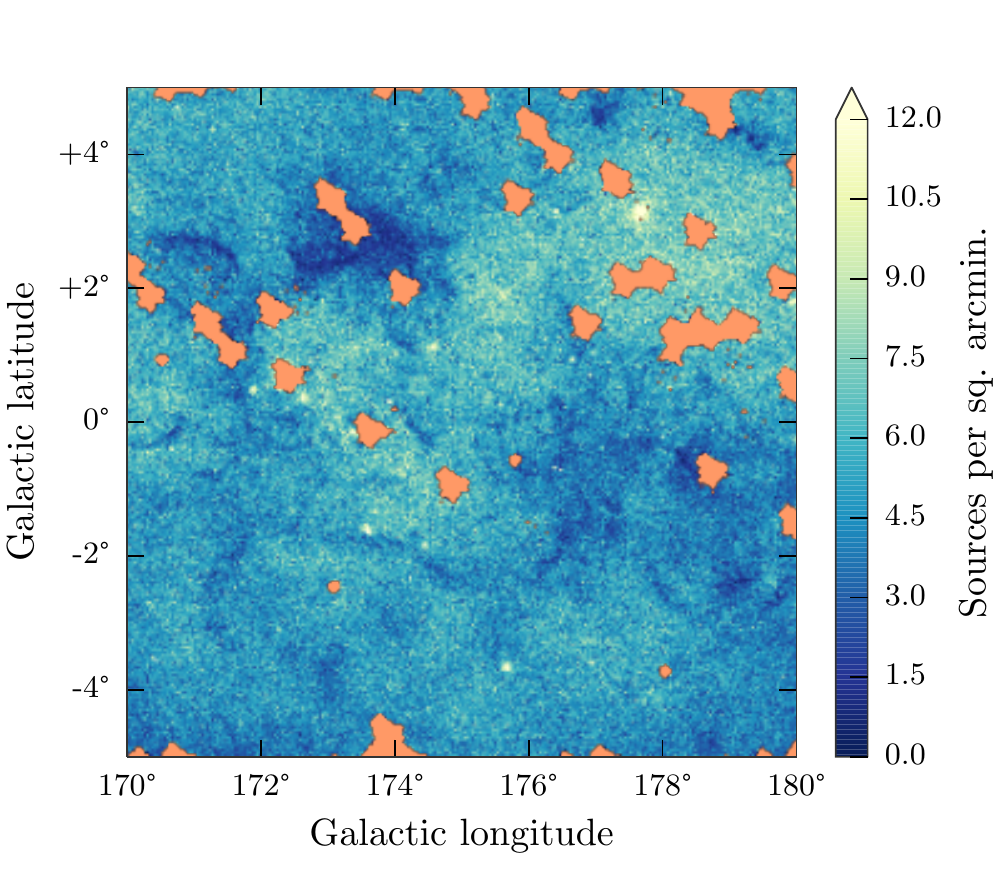} 
\caption{Cutouts of the $i<18$ density map with $2^{\prime}\times2^{\prime}$ resolution. Upper: Inner Galaxy sightline showing the Aquila Rift. Middle: Sightline covering the Cygnus region. Lower: Anticentre sightline.  The areas shaded in \textbf{salmon pink} are regions where no counts are presently available from the IPHAS DR2 catalogue, or where regions of 5$^{\prime\prime}$ have been masked around stars brighter than $V=5$.  Note that the colour scales applied in the different panels are different.}
\label{fig:dmap_cutout}
\end{center}
\end{figure}

\begin{figure*}
\includegraphics[width=1.0\linewidth]{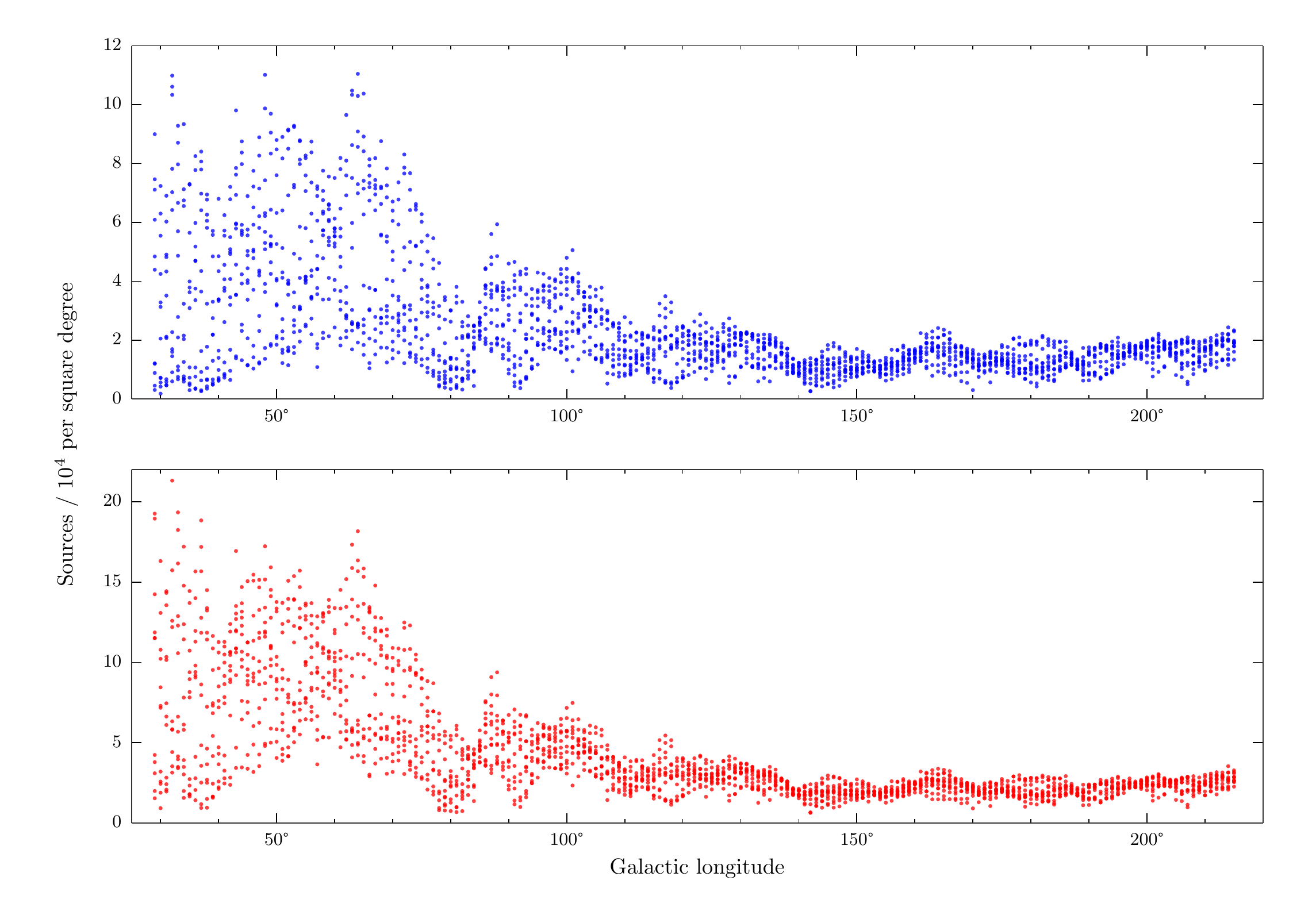}
\caption{Source counts for $r<19$ (\textbf{upper}) and $i<19$ (\textbf{lower}) as a function of Galactic longitude, per 1 sq. deg. cell (per-cell counts taken from the coarsest resolution density maps produced).}
\label{fig:dmap_both_bands}
\end{figure*}

Both $r$- and $i$-band density maps are available for a range of depths and resolutions. The resolutions available begin at $1^{\prime}\times1^{\prime}$ and increase to cells of $2^{\prime}$, $10^{\prime}$, $30^{\prime}$ and $1^{\circ}$ on a side. At each resolution the maps are stored as multi-extension FITS files, with the extensions containing maps of increasingly faint limiting magnitude. Table~\ref{table:fits_extensions} lists which extension corresponds to which faint limit for each band.  These files are provided as supplementary materials to this paper and they will also be accessible from CDS and from the IPHAS survey website\footnote{http://www.iphas.org/}.  In the Appendix, we show
plots of the full final stellar density maps at $2^{\prime}\times2^{\prime}$ resolution, for both photometric bands. These are accompanied by plots showing the scale of the incompleteness corrections that had to be applied.

For the purpose of immediate illustration, three cutouts at $2^{\prime}\times2^{\prime}$ arcmin$^2$, taken from the $i$-band density map to sample contrasting Galactic longitudes are presented in Fig.~\ref{fig:dmap_cutout}.  The inner Galaxy sightlines in the range $30^{\circ} < \ell < 40^{\circ}$ (top panel in the figure) contain some of the lowest and the highest density cells in the entire map, owing to the most extreme contrasts between high equatorial extinction and intrinsically high stellar densities away from the mid-plane.  These pronounced contrasts ease off with increasing longitude swinging around toward the outer disc, as seen in the the $80^{\circ}<\ell<90^{\circ}$ and Anticentre sightlines also presented in Fig.~\ref{fig:dmap_cutout}. The areas left grey in these cutouts are areas for which counts are not yet available: among these, the smaller circular patches will be zones of avoidance around bright stars, while the larger areas with a more ragged outline will be due to no data having been accepted into IPHAS DR2.  The latter are much more common in the Anticentre region that is overhead in the La Palma winter, when the weather is less clement on average. 

At resolutions on the arcminute scale the density maps betray the fractal nature of the interstellar medium, presenting many fine dark filaments due to obscuring dust - some of these are especially easily picked out in the middle panel of 
Fig.~\ref{fig:dmap_cutout}.  In experimental comparisons we have made, we have noticed that much of this detail echoes the fine structure seen in the CO maps of \citet{Dame2001} presented at a spatial resolution of 7.5 arcminutes and in the more recent Planck maps of Galactic dust at 5 arcminutes resolution.  Indeed the effective resolution in the star counts is higher -- we particularly recommend the $2^{\prime}\times2^{\prime}$ sampling, as the median count per cell is 25 and  benefit is gained from the effective interpolation over smaller gaps in survey coverage. 

Already known star clusters frequently stand out.  In the $30^{\circ} < \ell < 40^{\circ}$ region, the globular cluster NGC 6749 is apparent as a count peak at $\ell = 36^{\circ}.167$, $b = -2^{\circ}.178$.  In the vicinity of the Cygnus-X region shown in the $80^{\circ} < \ell < 90^{\circ}$ cut-out, the open clusters Berkeley~54 and NGC~7044 can be noticed as peaks at $\ell = 83^{\circ}.129$, $b = -4^{\circ}.143$ and at $\ell = 85^{\circ}.890$, $b = -4^{\circ}.150$ respectively.  In the generally quieter star counts background of the Anticentre region, star clusters become even more evident to inspection: M37 particularly stands out as a broad bright peak at Galactic coordinates, $177^{\circ}.64$, $+3^{\circ}.09$.  In addition, Czernik~21 ($171^{\circ}.89$, $+0^{\circ}.45$), NGC 1907 ($172^{\circ}.62$, $+0^{\circ}.31$), NGC 1893 ($173^{\circ}.59$, $-1^{\circ}.68$) and Berkeley~17 ($175^{\circ}.65$, $-3^{\circ}.65$) are all obvious to the eye.  A possibility to pursue in the future is to take advantage of the uniformity and red sensitivity of these star counts to eliminate 
{\it false} clusters in the literature: this can be achieved by cross-matching the maps with published cluster catalogues 
to confirm (or not) that clusters identified from blue images are also apparent in the $r$ and $i$ bands.

Fig.~\ref{fig:dmap_both_bands} shows the variation in source counts per square degree across the Galactic Plane, with counts taken directly from the $1^{\circ}\times1^{\circ}$ density map (the coarsest resolution produced). The plot is similar to fig. 3 from \citet{Gonzalez-Solares2008}, with some differences.  The older diagram was based simply on counting rows per $\sim$0.25 square degree survey field in the IPHAS Intermediate Data Release (IDR), whilst the new ones are the result of rigorously counting sources in each of the $r$ and $i$ photometric bands down to the specified calibrated magnitude limits.  The imposition of a magnitude limit naturally reduces the source count such that now the typical stellar density in the outer disc is in the vicinity of 20000 per square degree in both bands (the count being higher in $i$), as compared to 50000 per square degree.  Otherwise, the general pattern in the dependence of stellar density on Galactic longitude remains much the same, including some of the substructure (e.g. the peak at $\ell \simeq 87^{\circ}$).  As a general rule the stellar surface density in the $i$ band is between 1.5 and 2 times higher than in the $r$ band, thanks to reduced extinction.

\begin{table}
\centering
\begin{tabular}{|c|ccccccc|}
\hline
 & 17.0 & 17.5 & 18.0 & 18.5 & 19.0 & 19.5 & 20.0  \\
\hline
$r$ &	- &	- &	1 &	2 &	3 &	4 &	5 \\
$i$ &	1 &	2 &	3 &	4 &	5 &	- &	-  \\
\hline
\end{tabular}
\caption{The FITS extension corresponding to faint limiting magnitudes for both $r$- and $i$-band density maps.}
\label{table:fits_extensions}
\end{table}

\section[]{First comparisons with Galactic model predictions}
\label{sec:galactic_predictions}

In this final section of results, we provide a limited demonstration of how these star counts could be deployed as combined tests of Galactic models and 3D extinction maps.  It will certainly be appropriate to take such tests further, but these first comparison provides evidence that the recent extinction maps of \citet{Sale2014} are an improvement on the older 2MASS-based maps of Marshall etal (2006).

\subsection{Choice of Galactic model}

The Besan\c{c}on model \citep{Robin2003} has been available to the astronomical community for over a decade now: it is 
designed to generate synthetic Galactic populations, compiling knowledge of the Milky Way from several sources. The required inputs to the model include stellar density gradients, star formation history, age-metallicity relationships and a stellar initial mass function. It is one of the tools that will be used to interpret Gaia data, and is being updated to 
incorporate the latest developments in studies of Galactic evolution, structure and kinematics \citep{Czekaj2014}. 
However, since the older 2003 model has already been used extensively \citep{Ivezic2008,Momany2006} 
and continues to 
offer downloadable synthetic catalogues\footnote{http://model.obs-Besancon.fr/}, we use it here to make the very first comparisons against IPHAS completeness-corrected star counts.

The synthetic stellar catalogues delivered by the Besan\c{c}on website can be specified in terms of both
heliocentric distance and Galactic coordinates. The parametrization of the overlaid extinction is adjustable, 
while spectral types, along with absolute and apparent magnitude ranges, can also be specified. 
Two output photometric systems are available: Johnson-Cousins and CFHTLS-Megacam. For the present purpose of 
comparison with IPHAS data, the SDSS-like MegaCam filters are the appropriate choice. The IPHAS/SDSS 
transformations presented in \citet{Barentsen2014} have been applied here to bring the returned synthetic 
magnitudes into the IPHAS photometric system.  We have retrieved simulated catalogues for three 
$2^{\circ}\times10^{\circ}$ strips, perpendicular to the Galactic equatorial plane, in order to sample a 
range of likely outcomes (each one taken from a region plotted in Fig.~\ref{fig:dmap_cutout}).  
We have set the diffuse extinction parameter to zero in these requested simulations, so that we may correct 
the catalogues using more recently-published 3-D extinction data (see Section \ref{subsec:adding_extinction}). 
By limiting our attention to just these three strips, it became practical to collect the full range of stellar 
populations and to retain the default absolute magnitude range ($-7 < M < 20$). An apparent magnitude range of 
$8<r<25$ was imposed in our selections to exclude any objects falling far outside IPHAS detection 
limits.

While the Besan\c{c}on model adopts a thin disc with a hole at its centre alongside a separately specified bulge
component, we only need consider the thin disc component here. The influence of the central Galaxy specification can 
be neglected as even at the lowest Galactic longitudes covered by IPHAS, the sightlines towards the inner Galaxy have 
their closest approach $\sim$4 kpc from the Galactic Centre -- beyond the region affected by the bulge
component.

\subsection{The selected sightlines}

The regions chosen for comparison have been selected to come from contrasting longitudes ($\ell \approx
30^{\circ}$, $90^{\circ}$, and $175^{\circ}$), with the precise choices designed to avoid gaps in 
IPHAS DR2 coverage.  From the Besan\c{c}on web interface, we have extracted the relevant $i$-band catalogues
with no applied extinction.

\begin{figure*}
\begin{center}
\includegraphics[width=0.75\linewidth]{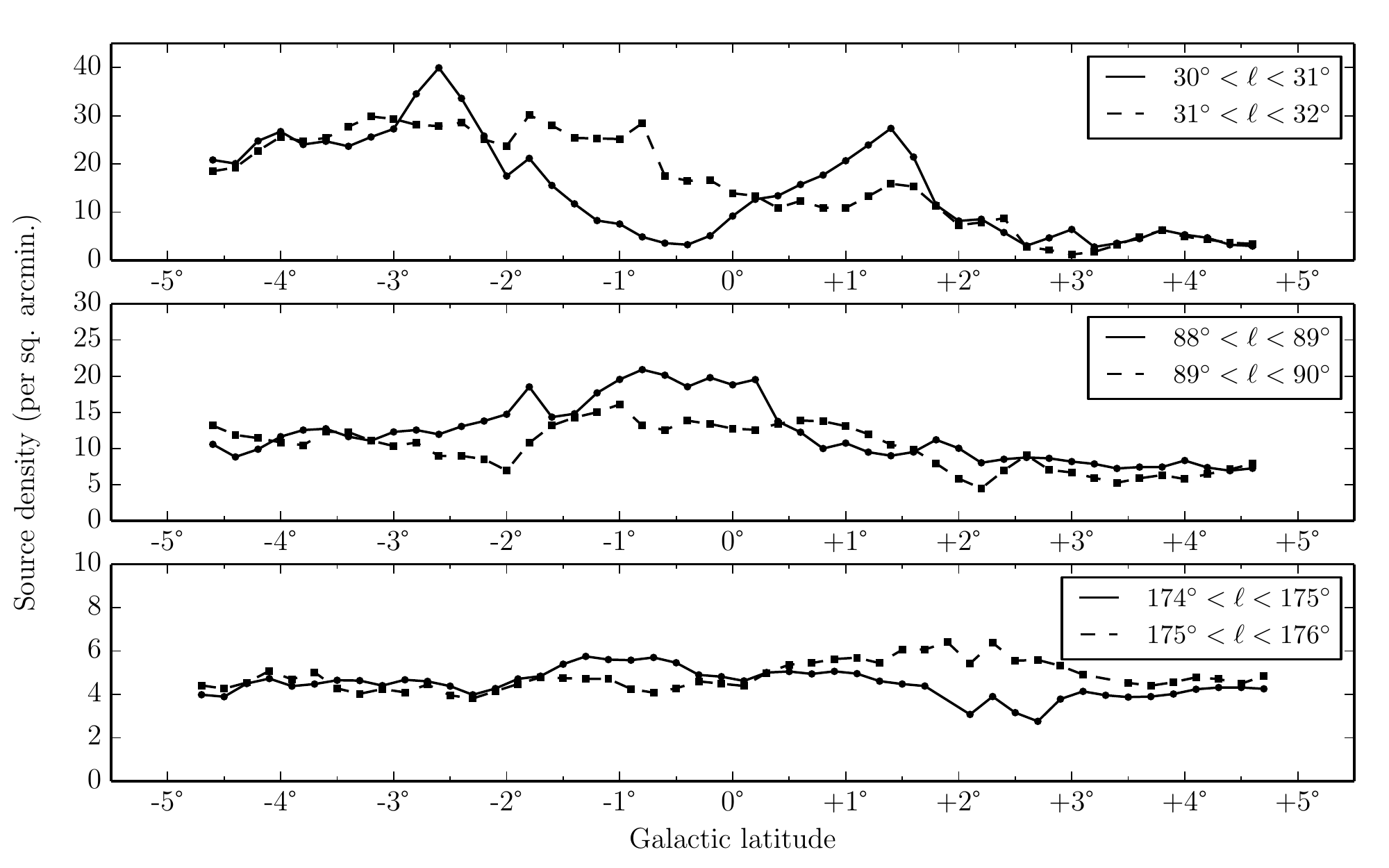} 
\caption{Stellar density profiles in Galactic latitude, averaged over $1^{\circ}$ wide strips in Galactic longitude, as given by the $20^{\prime}\times20^{\prime}$ resolution IPHAS $i$-band density map, for a magnitude limit of 18. Each panel shows two neighbouring strips for each of the three general directions within the 
Galactic Plane.}
\label{fig:iphas_densities}
\end{center}
\end{figure*}

Fig.~\ref{fig:iphas_densities} shows the variation in stellar densities with Galactic latitude for three pairs of strips, 
as retrieved from the $20^{\prime}\times20^{\prime}$ resolution IPHAS $i<18$ density map. At each of the three longitudes investigated, the star count data for two adjacent $1^{\circ}$-wide strips are shown to provide a more 
representative impression of the degree-scale variations present. 

\begin{figure}
\begin{center}
\includegraphics[width=1.0\linewidth]{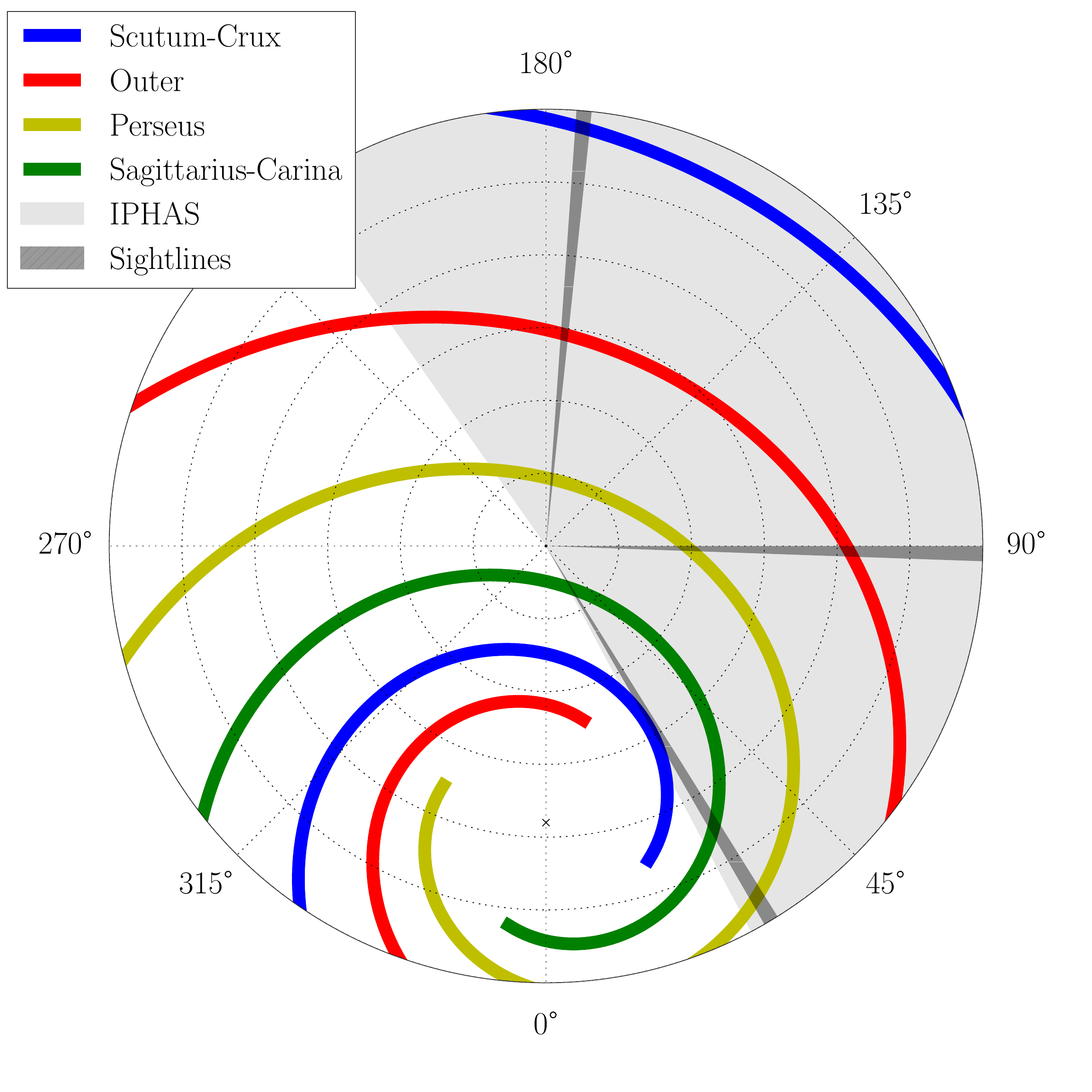}
\caption{Distribution of spiral arms in Galactic coordinates, with distance from the Sun to the spiral arms based on the parameters obtained by \citet{Vallee2008}. Dotted circles denote distances in increments of 2 kpc. The Galactic Centre is shown as a black cross. The region covered by IPHAS is highlighted in grey. The sightlines compared with Besan\c{c}on model predictions are denoted by darker hatched regions.}
\label{fig:spiral_arms}
\end{center}
\end{figure}

The inner Galaxy sightline shows the largest variation with latitude and the largest variation between the two separated 
$1^{\circ}$-wide strips. The Aquila region ($\ell\lesssim 45^{\circ}$) contains both the highest and lowest valued cells 
of the entire IPHAS density map. A dark cloud at ($\approx 30^{\circ}$, $-1^{\circ}$) is responsible for the reduction 
in stellar density (of the amplitude $\approx 10-20$ sources per sq. arcmin.) in the $30^{\circ}<\ell <31^{\circ}$ profile 
relative to the $31^{\circ}<\ell <32^{\circ}$  profile. The Aquila Rift, just two to three hundred parsecs away, is responsible for the drop-off in stellar density at higher latitudes in both strips. Taking the spiral arm positions of \citet{Vallee2008}, it can be seen (in Fig.~\ref{fig:spiral_arms}) that this sightline will pass through the 
Sagittarius-Carina arm twice by a distance of 10 kpc, as well as follow the tangent of the Scutum-Crux arm: in 
combination with the local dust, such spiral arm crossings would be expected to produce the complex extinction 
distribution associated with this direction.

The sightline towards $\ell=90^{\circ}$, bordering the Cygnus region ($60^{\circ}\lesssim\ell\lesssim 90^{\circ}$), shows 
similarly pronounced variation with latitude, and between the two strips. Further from the Galactic midplane 
the two seem to be in good agreement. The spiral arm map of \citet{Vallee2008} places the Perseus arm at 
$\approx4$ kpc in this direction.  In contrast, the sightline passing close to the Anticentre shows the least variation: 
at positive latitudes the difference between the two adjacent strips is no more than $\approx 2$ per sq. 
arcmin. The flatness of these distributions highlights the relative lack of structure apparent in 
the density map at these longitudes; this suggests that the Besan\c{c}on model that does not attempt to emulate 
spiral arm structure is likely to perform well along this sightline -- a simple exponentially-decreasing stellar density with Galactocentric radius may capture the observed smooth behaviour. Nevertheless the Perseus Arm, described by 
\citet{Vallee2008} as lying at a distance of $\approx2$ kpc at this longitude, also passes through.  

\subsection{Adding in extinction to the synthetic catalogues}
\label{subsec:adding_extinction}
With the diffuse extinction parameter set to 0 mag kpc$^{-1}$, the synthetic catalogues returned provide an unreddened 
view of the Milky Way. As the catalogues specify the distance to each generated object, a custom extinction profile can 
be imposed on the simulated data. This has come from one of two empirically-based sources, as detailed below.

\subsubsection{Marshall et al extinction map}
The 3-D extinction map produced by \citet{Marshall2006} has been applied to the catalogues, taking the sightline closest 
to the synthetic catalogue coordinates - the \citet{Marshall2006} sightlines are binned in 0.25$^{\circ}$ increments in 
Galactic longitude and latitude. This map only covers sightlines with $\ell<100^{\circ}$, which means that the Anticentre
comparison we make cannot benefit from this model. It is based on analysis of 2MASS near-infrared photometry in 
combination with constraints from the Besan\c{c}on model: accordingly, the extinction in this case is specified for the 
$K_s$ band  ($A_{K_{s}}$) as a function of distance.  For present purposes, this requires the intermediate step of
conversion to $A_{i}$.  To do this we use extinction law data from \citet{Cardelli1989}, along with conversion 
factors for the K and I filters, thereby obtaining $A_{i}=4.2A_{K_{s}}$.

\subsubsection{Sale et al extinction map}
\label{sec:Sale-3D}
The 3D map of \citet{Sale2014} is the result of combining IPHAS DR2 photometry with the hierarchical Bayesian-inference model developed by \citet{Sale2012a}, which estimates the distance-extinction relationship along a given sightline, along with 
estimates of the atmospheric parameters of the stars sampled. A prior adopted in this treatment is that all sightlines are occupied 
only by thin disc stars, fitting to an exponential scale length of 3 kpc.  It is important to note that, although this 
extinction map was also constructed from IPHAS DR2, it is only the precision on the returned 
extinction to any given distance that carries a strong dependence on the stellar density distribution deduced from the photometric data in common.  In other words there is no circularity of argument here, as the comparisons below make plain.

The map provides a typical angular resolution of 10$^{\prime}$; the nearest sightline to the Besan\c{c}on catalogue under 
consideration was adopted in extinguishing the synthetic photometry. While each sightline comes with a warning of a maximum reliable distance (after which a prior on extinction takes over), holding the extinction constant beyond these limits would risk too large a contribution to the predicted star count from these distant regions.  For this reason, the extinction was 
permitted to continue to rise, even if it is no longer directly empirically constrained. All three longitudes considered here fall into the area treated by \citet{Sale2014}. 

\subsection{The observed and predicted star-count profiles compared} 
\label{sec:count_comparison}

\begin{figure*}
\begin{center}
\begin{flushleft}
\large{$\;\;\;\;\;\;\;\;\;\;\;\; \ell$=30$^{\circ}$:}
\end{flushleft}

\includegraphics[width=0.6\textwidth]{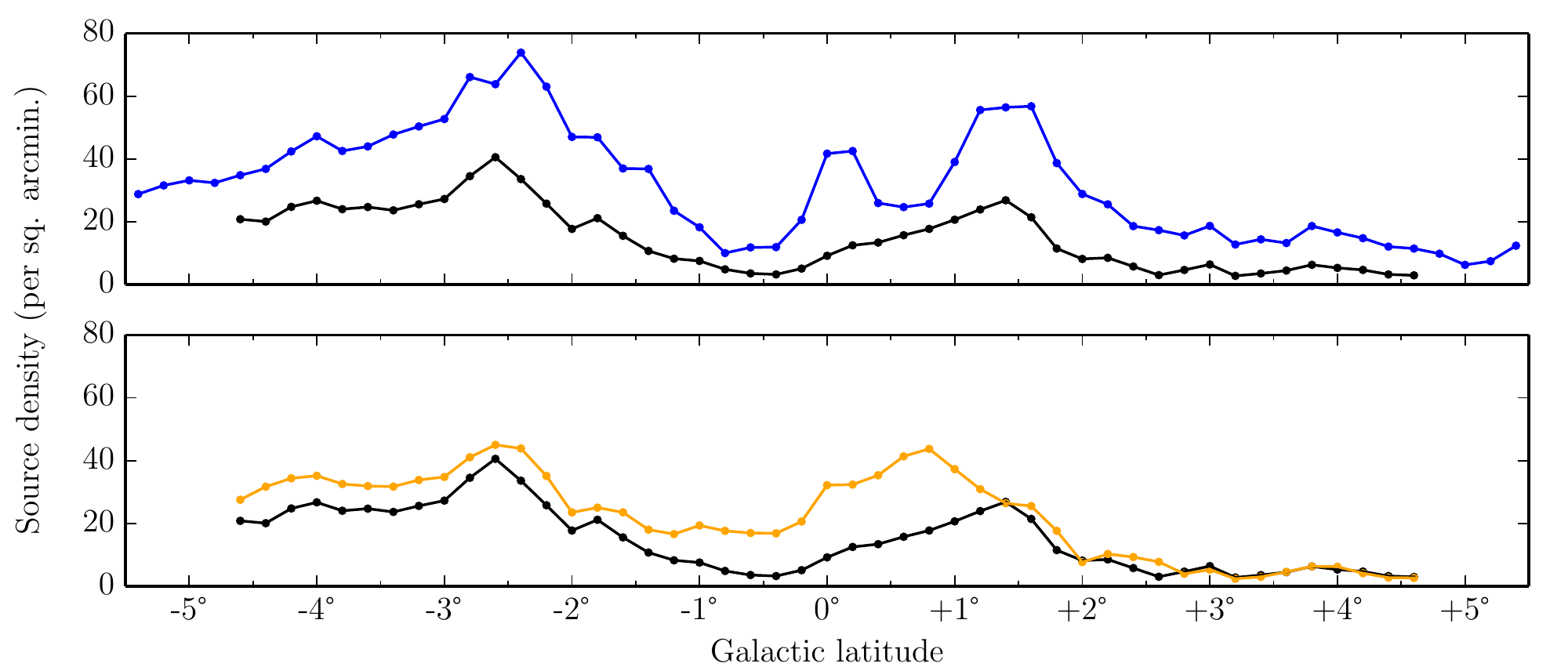} 

\begin{flushleft}
\large{$\;\;\;\;\;\;\;\;\;\;\;\; \ell$=90$^{\circ}$:}
\end{flushleft}

\includegraphics[width=0.6\textwidth]{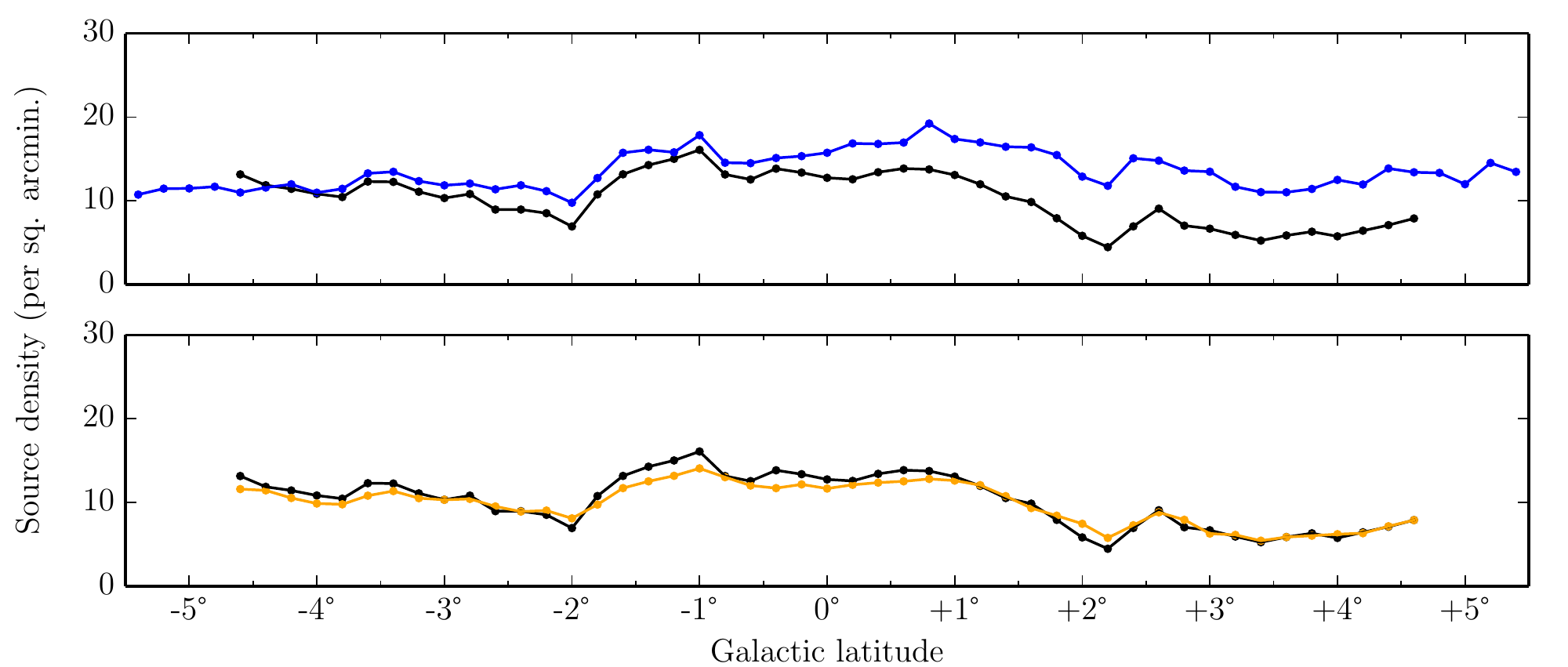}

\begin{flushleft}
\large{$\;\;\;\;\;\;\;\;\;\;\;\; \ell$=175$^{\circ}$:}
\end{flushleft}

\includegraphics[width=0.6\textwidth]{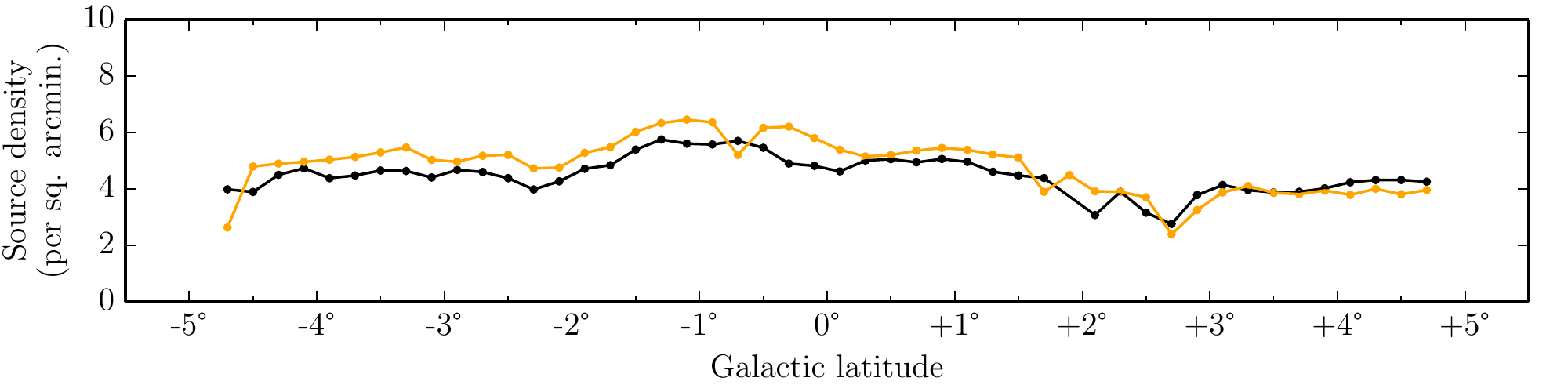} 
\caption{Comparisons between incompleteness-corrected IPHAS star counts to $i = 18$ (black) and Besan\c{c}on predictions, reddened by the extinction curves of \citet{Marshall2006} (blue) and \citet{Sale2014} (yellow). From top to bottom, comparisons are for sight lines at $\ell$=30$^{\circ}$, 90$^{\circ}$ and 175$^{\circ}$.  The errors in the empirical star-count data are 3--4 percent. }
\label{fig:count_comparison}
\end{center}
\end{figure*}

At each longitude, we make a comparison for just one of the two one-degree wide cross cuts at constant longitude.  These are shown in Fig.~\ref{fig:count_comparison}.  The strip sampling $\ell = 30^{\circ}$ to $31^{\circ}$ is compared to model predictions using both 3-D extinction maps.  It is striking that, irrespective of the extinction map applied, there is a marked over-prediction of the star count -- particularly near the Galactic Equator, where the discrepancies are on the scale of a factor two or more.  However, the \citet{Sale2014} extinction data do a somewhat better job than the $K_s$-band based \citet{Marshall2006} data, actually managing to agree at positive Galactic latitudes.  The need to convert $K_s$ band extinction to suit $i$ may partly be behind the weaker performance of the \citet{Marshall2006} modelling, but the differences in the `predicted' profile shapes also indicate that the two extinction maps describe different line-of-sight distributions of the dust.  One obvious difference is that \citet{Sale2014} mapping is better able to place the sharp rise in extinction due to the Aquila Rift, presenting anything from 3 to 7 magnitudes of extinction, within the first kiloparsec where it belongs.  Nevertheless, the biggest discrepancies around the Galactic equator coincide with the domain in which the maximum extinctions inferred in this mapping fall well short of the large total extinctions described in the \citet{Schlegel1998} map (see fig. 10 in \citet{Sale2014}), raising the possibility of too little in the 3D map.  There is a known bias of this sense that affects extinction mapping and it may have greater impact along these especially extinguished sightlines.

The second and third comparisons in Fig.~\ref{fig:count_comparison} at $\ell \simeq 90^{\circ}$ and  $\ell \simeq 175^{\circ}$ respectively use the $\ell = 89^{\circ}$ to $90^{\circ}$ and the $\ell = 174^{\circ}$ to $175^{\circ}$ strips presented in Fig.~\ref{fig:iphas_densities}.  Here the comparisons are encouraging, especially in the case of the \citet{Sale2014} extinction data, where the discrepancies are limited to under $\sim$one star per square arcminute in both cross cuts.  Certainly the agreement achieved at $\ell = 90^{\circ}$ is within likely error.

The less compelling performance at $\ell = 30^{\circ}$ compared to the agreement at the larger longitudes supports the assertion in section~\ref{sec:Sale-3D} that the shared IPHAS origin of the \citet{Sale2014} extinction data and the star counts does {\it not} guarantee a good match.  There can be two causes for the divergence between them --  either the dependence of extinction on distance is incorrectly represented (already considered above), or the mock star catalogue created from the Besan\c{c}on model is too well-populated.  It is not our aim in this limited discussion, to try to conclude on the origin of the discrepancyy as this would require the examination of a much more comprehensive set of comparisons.  Suffice it to say that, in the light of the good behaviour at $\ell \simeq 90^{\circ}$ and at $\ell = 175^{\circ}$, it would be premature to insist the cause is all in the 3D extinction mapping.  

For now, the possibility remains open that the problem is linked to the distinctive character of this inner Galaxy sightline: climbing up the stellar density gradient, it captures large numbers of more intrinsically-luminous highly-extinguished stars distributed along the pencil beam.  Interrogation of the mock catalogues shows us that the median distance to objects at $\ell \simeq 30^{\circ}$ included in the $i < 18$ star counts is 5--6 kpc, as compared with 2--3 kpc at $\ell \simeq 175^{\circ}$. In other words, at $\ell = 30^{\circ}$, the IPHAS DR2 star counts are, of necessity,  made up from more luminous stars than those at $\ell = 90^{\circ}$ and $\ell = 175^{\circ}$.   Too many intrinsically brighter stars is equivalent to a stellar luminosity function too well-populated at the high end, and hence may imply a star formation history weighted too strongly to recent times.  \citet{Czekaj2014} have encountered a similar difficulty in comparing the Tycho-2 catalogue (limited to $V < 11$, and the Solar neighbourhood) with predictions based on an updated version of the Besan\c{c}on model: they also predict too high a star count in the same part of the Galactic Plane  here (see their fig. 15) -- even after imposing a declining star formation rate over the past 10 Gyrs.  Here, the comparison is with a Galactic model assuming constant star formation.  It will be interesting to examine these issues more closely, taking advantage of the much larger Galactic disc volume these newly-provided fainter IPHAS counts can probe. 

\section{Concluding remarks}

Our main goal has been to present new IPHAS-based faint $r$ and $i$ star counts for the northern Galactic Plane, the first deep optical counts for the northern Milky Way.  In view of the varying observing conditions that have fed into the IPHAS DR2 database, it has been important to correct the derived stellar density map for incompleteness due both to confusion and and to declining sensitivity near the faint limit on the angular scale that accounts appropriately for varying observing conditions.  Here, that scale is set by the camera footprint of 
$\sim$0.29 square degrees.  

Although it is relatively computation-intensive to use artificial source injection to determine these corrections, our presentation of the alternatives confirms that it is the best available route to placing these faint star counts within a precise quantitative framework.  At the same time, the more approximate methods provide insight and supporting evidence that incompleteness at Vega magnitudes brighter than $i = 18$ and $r = 19$ is small. Application of the preferred artificial source method across 94\% of the IPHAS 1800 square-degree footprint, has traced the appearance of the Galactic disk down to angular resolutions as fine as 1 square arcminute.  The $i$ band counts are commended for use down to 19th magnitude, and those in $r$ down to 20th.  At these limits the computed completeness fractions are generally above 0.8.

We described in Section~\ref{sec:the_maps} how the stellar density data are available for re-use in the form of digital maps to a range of limiting magnitudes and at a range of angular resolutions.  These are provided with this paper as supplementary material and also may be downloaded from the survey website (see http://www/iphas.org/data/).  

There is a limitation to point out.  The method discussed in Section \ref{sec:completeness} takes the average completeness of the central CCD of the WFC as being representative of the complete (4-CCD) IPHAS field. This takes proper account of the prevailing observing conditions, but does not trace local density variations due to e.g. compact star clusters. In regions with small average fractional corrections (under 10 percent), the corrected density maps will nevertheless remain accurate even on the $\sim$arcminute scale, since the directly measured count will still be reliable. In contrast, in the minority of sky regions with higher fractional corrections and strong stellar density variations within the WFC footprint, under-estimation is liable to affect comparisons of star counts at scales below $\sim10^{\prime}\times10^{\prime}$. Putting this point another way, the star counts presented here should not be assumed to yield accurate peak stellar densities in compact inner-Galaxy clusters.  At the same time, the higher resolution maps are well-suited to the uniform detection of star clusters, creating an opportunity to re-evaluate the content of current optically-based cluster catalogues.

Over the next decade Gaia photometry based on the mission's customised system of bandpasses, will no doubt become the basis for important work in this area.  In this context, it is useful to note that broadband Gaia $G \simeq 20$ roughly corresponds to IPHAS $i \simeq 19$ (Vega) for the reddened populations dominating the faint star counts, while a similar cross-comparison of the magnitude scales for the Gaia blue and red prisms indicates $G_{BP} \simeq 20$ matching to IPHAS $r \simeq 19$.  The $G_{RP}$ and IPHAS $i$ scales are similar.  These transformations are considered in more detail by \citet{Farnhill2015} and are based on the work of \citet{Jordi2006}.  The star counts presented here thus reach to the same or slightly greater depth as those likely to ultimately emerge as the Gaia mission completes. The important difference is that to brighter magnitude limits, out to a distance of $\sim$ 1 kpc, Gaia counts will benefit from the high angular resolution achieved in space that -- together with the advanced astrometric capability -- will expose 60\% of binaries to 250~pc and 35\% of binaries to $\sim$1~kpc\footnote{http://sci.esa.int/gaia/31441-binary-stars/} (a star like the Sun will be $r \sim 16$ at 1 kpc). An exhaustive accounting of this volume will be achieved, and stellar multiplicities will be exposed. These specific advantages decay away on the one-to-a-few kiloparsec scale to which the source counts presented here are most sensitive.  

To illustrate the potential application of the IPHAS maps, we have presented a first example of comparison between the measured $i$ counts and the predictions of the \citet{Robin2003} Galactic model (Section~\ref{sec:count_comparison}). These star counts, down to 18th magnitude, are capable of sampling the Galactic disc well beyond the Solar neighbourhood, offering a combined test of inferences on the 3D distribution of interstellar dust and of assumptions shaping the Galactic stellar luminosity function.  In these first limited tests, the 3D extinction map of \citet{Sale2014} appears to be faring very well along sightlines passing through larger Galactocentric radii, while first signs of a challenge appear for sightlines plunging inwards of the Solar Circle.  The origin of the emerging differences deserve further closer examination, that can draw on a wider selection of count comparisons and also on the information encoded within IPHAS DR2 colour-magnitude data. This represents a particularly appropriate use of IPHAS survey photometry in that both the available Galactic models and these star counts best describe behaviours on angular scales of 0.2 to 1 degree, averaging over individual clouds and clusters. In these angular and depth domains, these new IPHAS-based stellar density maps are ready to play a validating role, now, in the quest to map out the relative 3D distributions of interstellar dust and the stars in the Galactic plane. Work toward a further paper applying these new counts in a broad examination of the predictions of current Galactic 
models is now underway.

\section*{Acknowledgements}

This paper makes use of data obtained as part of the INT Photometric H$\alpha$ Survey of the Northern Galactic Plane (IPHAS, www.iphas.org) carried out at the Isaac Newton Telescope (INT). The INT is operated on the island of La Palma by the Isaac Newton Group in the Spanish Observatorio del Roque de los Muchachos of the Instituto de Astrofisica de Canarias. All IPHAS data are processed by the Cambridge Astronomical Survey Unit, at the Institute of Astronomy in Cambridge.

The bandmerged DR2 catalogue underpinning this work was assembled at the Centre for Astrophysics Research, University of Hertfordshire, supported by a grant awarded by the Science \& Technology Facilities Council (STFC) of the United Kingdom [ST/J001333/1]. HJF also acknowledges the receipt of a PhD studentship funded by the STFC.

This work made use of the \textsc{topcat} \citep{Taylor2005} and \textsc{astropy} \citep{Astropy2013} packages.

The authors thank Stuart Sale for useful conversations related to the first exploratory test of Galactic modelling presented here, and Yvonne Unruh for comments more broadly on the text.  An anonymous referee is also thanked for comments that have also led to improvements of this paper's content.

\bibliographystyle{mn2e}
\bibliography{dmap}

\appendix

\section{Plots of the $2^{\prime}\times2^{\prime}$ density maps}
\label{sec:full_plots}

We present plots of the $r$ and $i$ density maps, down to limiting magnitudes of 19th and 18th respectively (in the Vega system) and at a resolution of $2\times 2$ arcmin$^2$.  The two maps are broken into 30-degree sections, and are shown alongside the corresponding maps of fractional completeness correction 
so as to highlight where high stellar densities, sometimes in combination with worse-than-median seeing, lead to a need for larger corrections.  The reader is reminded that median seeing is 1.1 arcsec in $r$, and 1.0 arcsec in $i$ (see \citet{Barentsen2014}).  These plots also show where there are gaps in the coverage of the northern Galactic plane: the overall fraction of the northern plane covered is 0.94, with more gaps appearing in the outer disk. How the coverage changes with Galactic longitude is laid out in Table~\ref{table:coverage_fraction}.

\begin{table}
\begin{center}
\begin{tabular}{|cc|}
\hline
Longitude Section & Coverage \\
\hline
$30^{\circ}<\ell<60^{\circ}$ & 96.7\% \\
$60^{\circ}<\ell<90^{\circ}$ & 98.4\% \\
$90^{\circ}<\ell<120^{\circ}$ & 98.1\% \\
$120^{\circ}<\ell<150^{\circ}$ & 92.0\% \\
$150^{\circ}<\ell<180^{\circ}$ & 94.4\% \\
$180^{\circ}<\ell<210^{\circ}$ & 90.3\% \\
\hline
\end{tabular}
\caption{Coverage fraction for each segment of the density maps as shown in Appendix \ref{sec:full_plots}, inside latitude range $-5^{\circ}<b<+5^{\circ}$}
\label{table:coverage_fraction}
\end{center}
\end{table}

\begin{figure*}
\begin{center}
\includegraphics[width=1.0\textwidth]{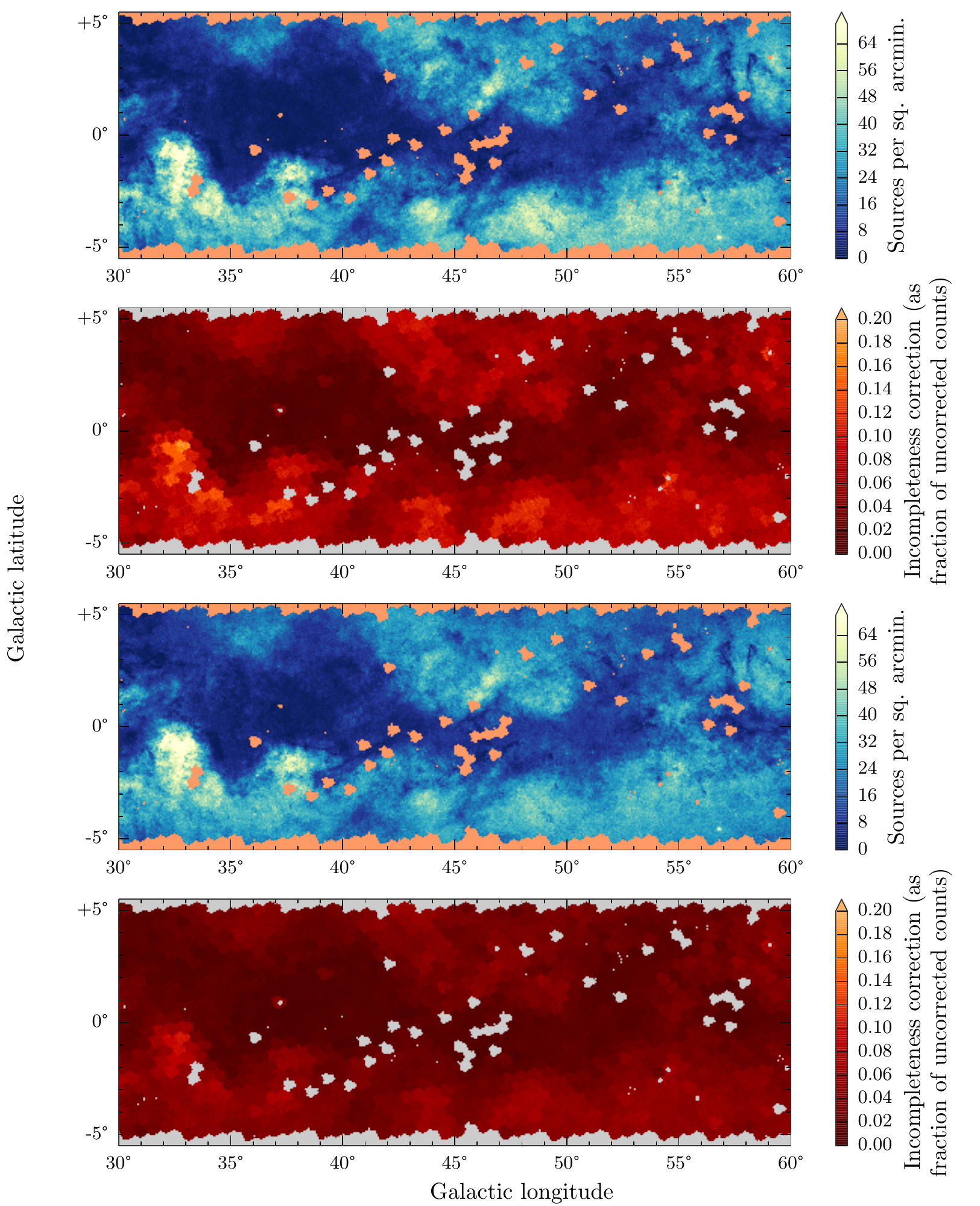} 
\caption{Cutouts of the $2^{\prime}\times2^{\prime}$ resolution density maps for the region $30^{\circ}<\ell<60^{\circ}$. \textbf{a)} $r$-band density map of sources down to 19th mag, \textbf{b)} $r$-band corrections down to 19th mag, \textbf{c)} $i$-band density map down to 18th mag, \textbf{d)} $i$-band corrections down to 18th mag.  \textbf{In (a) and (c), the areas shaded salmon pink are locations without included data presently.}}
\label{fig:dmaps_30}
\end{center}
\end{figure*}

\begin{figure*}
\begin{center}
\includegraphics[width=1.0\textwidth]{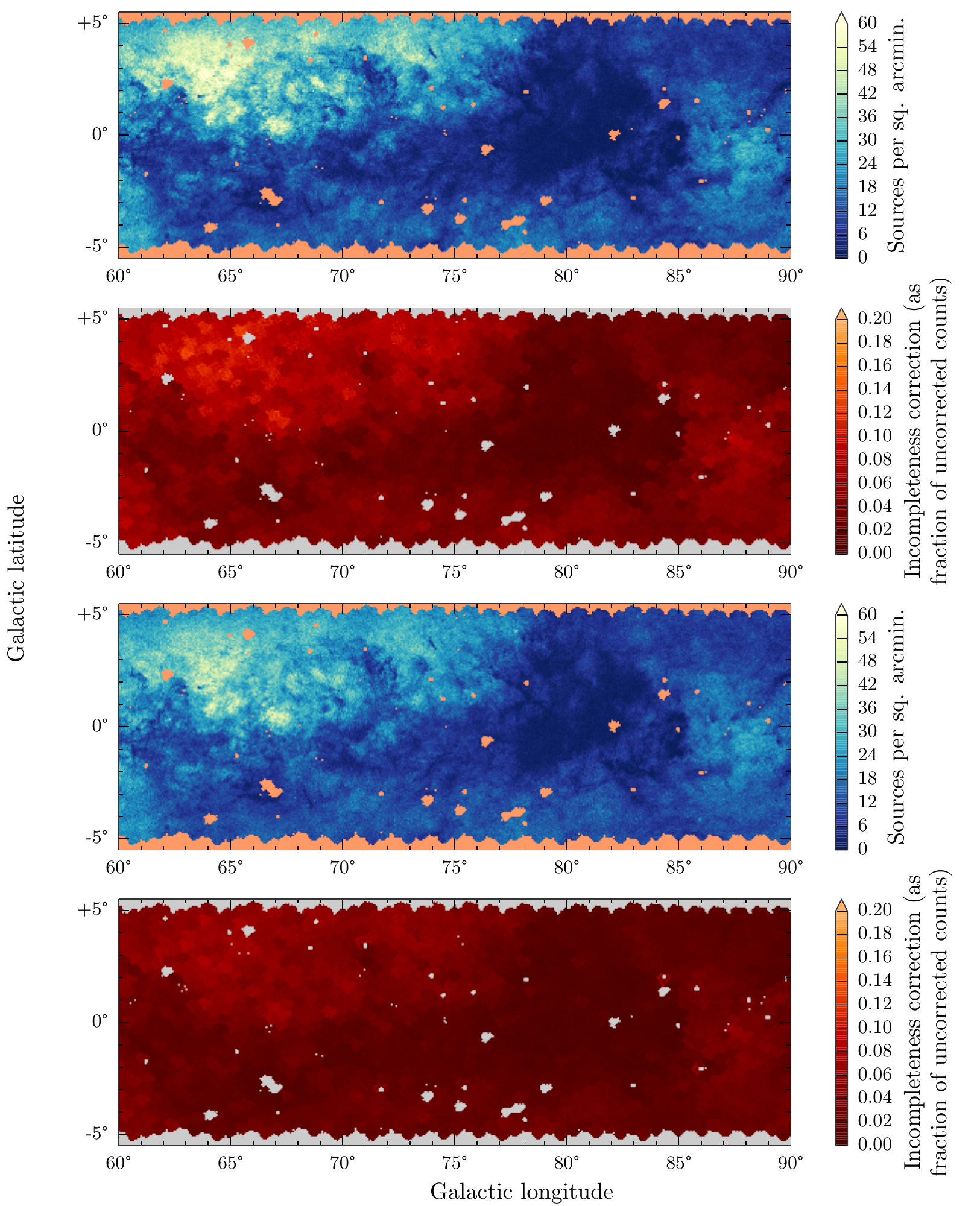} 
\caption{As Fig.~\ref{fig:dmaps_30}, only for $60^{\circ}<\ell<90^{\circ}$.}
\label{fig:dmaps_60}
\end{center}
\end{figure*}

\begin{figure*}
\begin{center}
\includegraphics[width=1.0\textwidth]{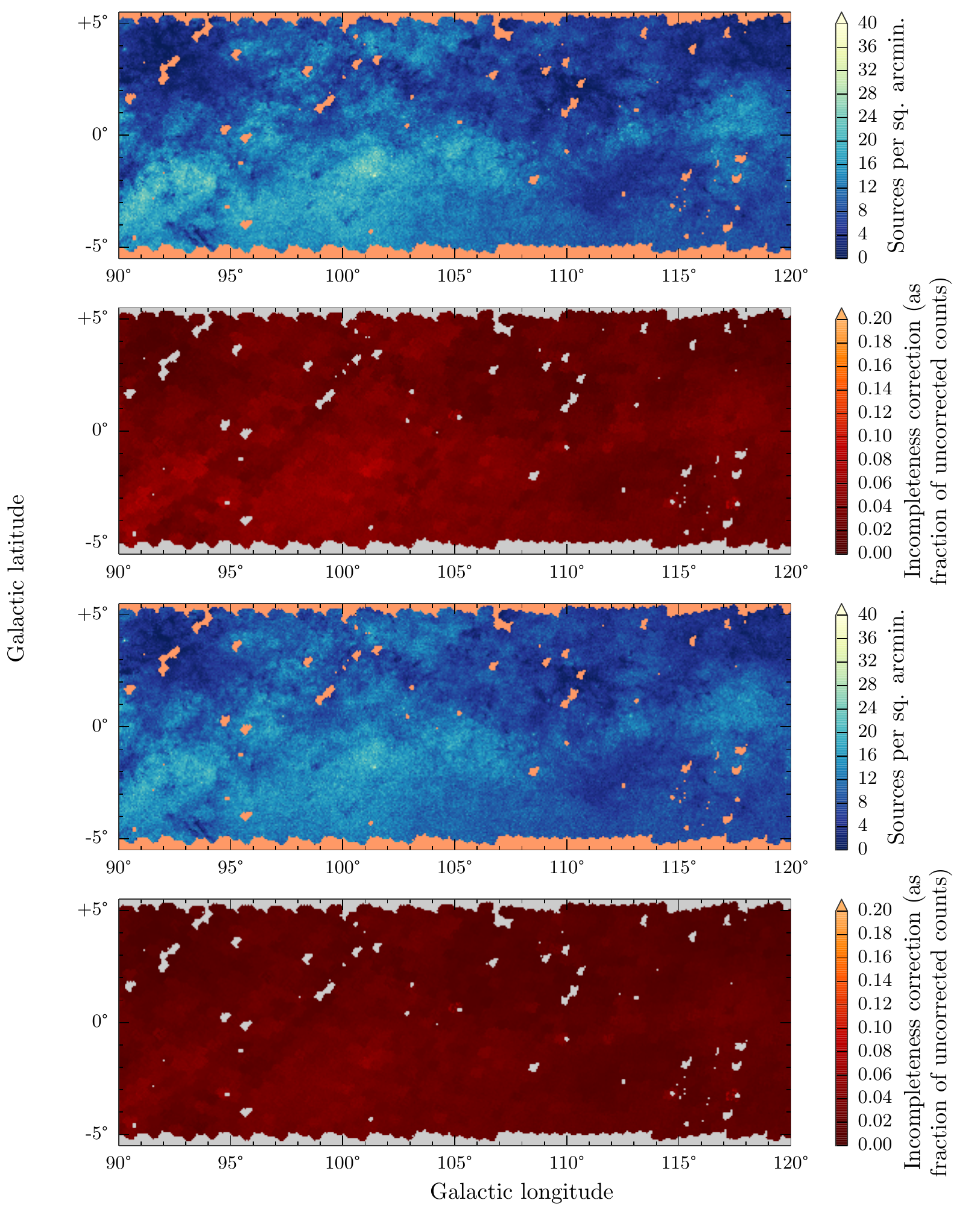} 
\caption{As Fig.~\ref{fig:dmaps_30}, only for $90^{\circ}<\ell<120^{\circ}$.}

\label{fig:dmaps_90}
\end{center}
\end{figure*}

\begin{figure*}
\begin{center}
\includegraphics[width=1.0\textwidth]{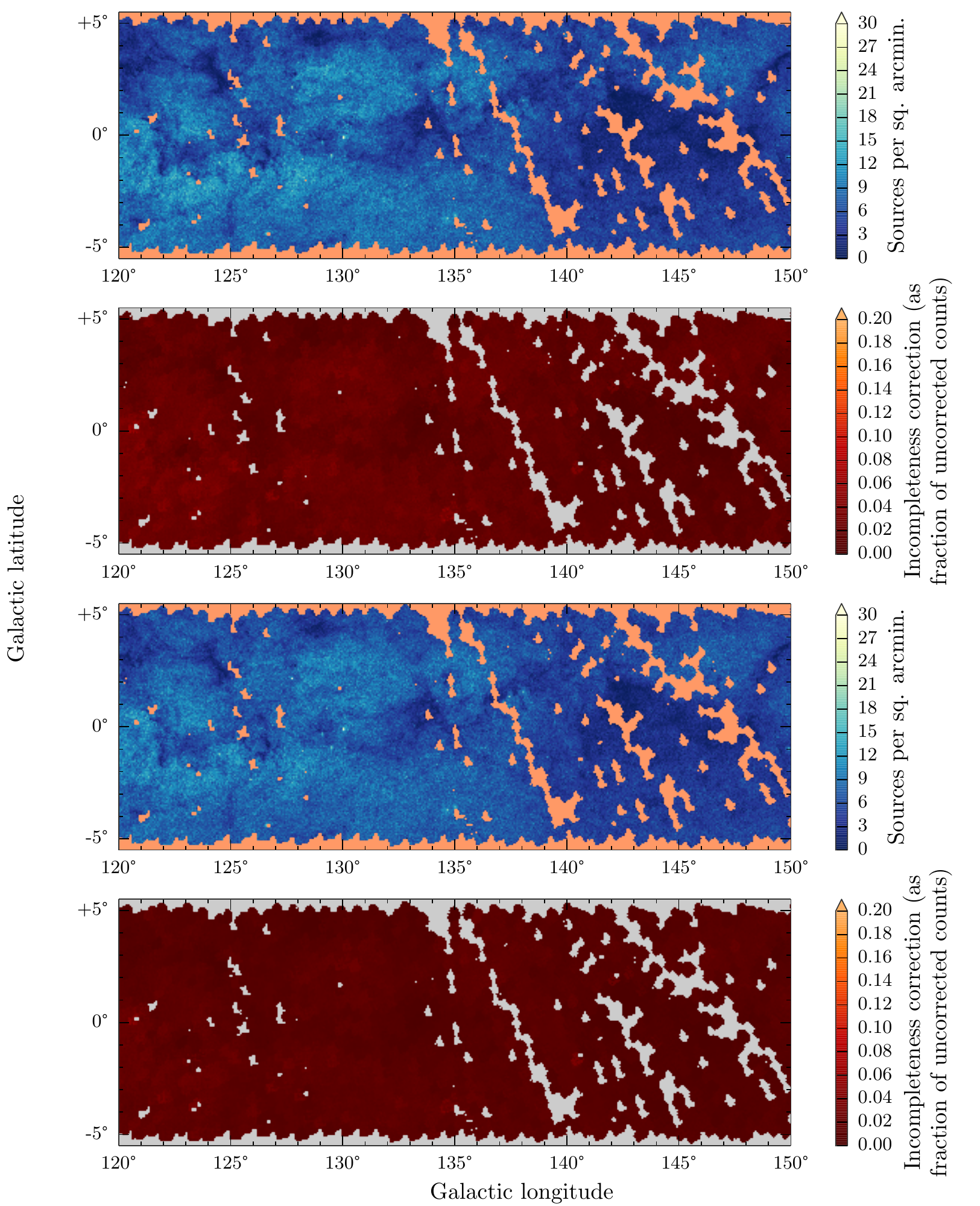} 
\caption{As Fig.~\ref{fig:dmaps_30}, only for $120^{\circ}<\ell<150^{\circ}$.}

\label{fig:dmaps_120}
\end{center}
\end{figure*}

\begin{figure*}
\begin{center}
\includegraphics[width=1.0\textwidth]{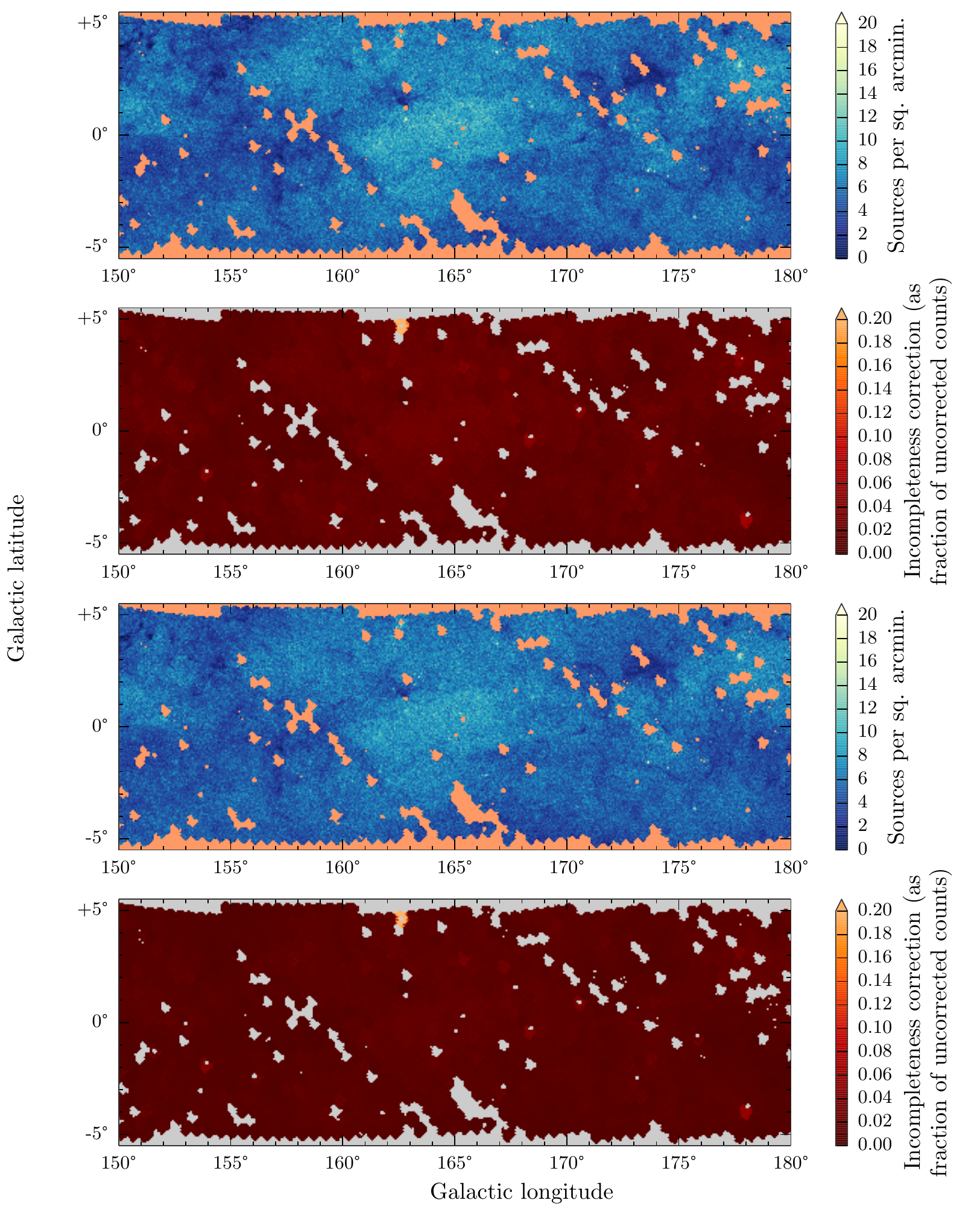} 
\caption{As Fig.~\ref{fig:dmaps_30}, only for $150^{\circ}<\ell<180^{\circ}$.}

\label{fig:dmaps_150}
\end{center}
\end{figure*}

\begin{figure*}
\begin{center}
\includegraphics[width=1.0\textwidth]{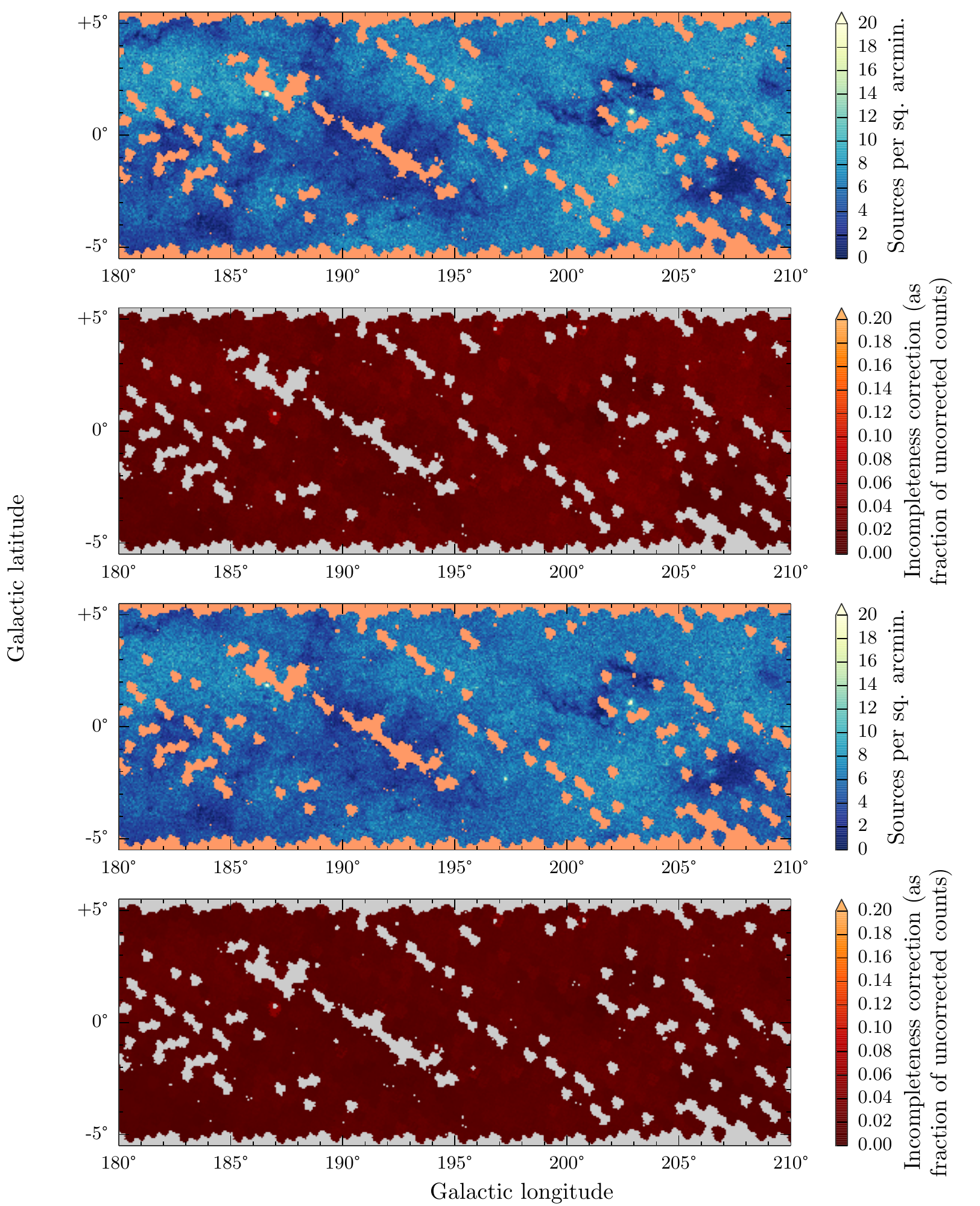} 
\caption{As Fig.~\ref{fig:dmaps_30}, only for $180^{\circ}<\ell<210^{\circ}$.}

\label{fig:dmaps_180}
\end{center}
\end{figure*}

\label{lastpage}

\end{document}